\documentclass[twocolumn,preprintnumbers,amsmath,amssymb]{revtex4}

\usepackage{graphicx}
\usepackage{latexsym}
\usepackage{amsmath,amscd}
\usepackage{graphics}
\usepackage{amssymb}
\usepackage{layout}
\usepackage{amsfonts,epsfig}
\usepackage{csquotes}
\usepackage{bm}
\usepackage{bbm}
\usepackage{comment}
\usepackage{outlines}
\usepackage{enumitem}
\usepackage{braket}
\usepackage{bbold}
\usepackage{pifont}
\usepackage{stackengine}
\setstackgap{S}{1pt}

\usepackage[usenames,dvipsnames]{color}

\begin{document}

\title{Topological insulator ring with magnetic impurities}
\author{Arian Vezvaee} 
\author{Antonio Russo} 
\author{Sophia E. Economou}
\author{Edwin Barnes}\email{efbarnes@vt.edu}
\affiliation{Department of Physics, Virginia Tech, Blacksburg, Virginia 24061, USA}

\begin{abstract}
Topological insulators exhibit gapless edge or surface states that are topologically protected by time-reversal symmetry. However, several promising candidates for topologically insulating materials (such as Bi$_2$Se$_3$ and HgTe) contain spinful nuclei or other types of magnetic impurities that break time-reversal symmetry. We investigate the consequences of such impurities coupled to electronic edge states in a topological insulator quantum ring threaded by a magnetic flux. We use spin conservation and additional symmetry arguments to derive a universal formula for the spectrum of propagating edge modes in terms of the amplitude of transmission through the impurity. Our results apply for impurities of arbitrary spin. We show that there exists an energy regime in which the spectrum becomes nearly independent of the flux and significant spectral gaps form. We further analyze the electron-impurity entanglement entropy, finding that maximal entanglement occurs near the gaps in the spectrum. Our predictions can be investigated with quantum ring transport interference experiments or through spin-resolved STM measurements, providing a new approach to understand the role of impurities in topological insulator edge transport.
\end{abstract}
\maketitle

\section{Introduction and motivation}

Topological insulators (TIs) are an interesting class of materials that behave as insulators in the bulk while exhibiting conducting helical surface or edge states~\cite{Qi/Zhang,Kane2005-1,Kane2005-2}. TIs are invariant under time-reversal (TR) symmetry, and their surface or edge states are topologically protected provided this symmetry remains unbroken \cite{Kane2005-1,Moore,Roushan}. These states have spin and momentum locked orthogonally to each other, and hence states of opposite momentum have opposite spin so that full backscattering cannot occur without a spin-flipping mechanism~\cite{Moore}. One of the best known examples of TIs are HgTe quantum wells, first predicted by Bernevig, Hughes and Zhang (BHZ) ~\cite{BHZ} and later confirmed in various experiments~\cite{Konig2007,Roth,Konig2008,Buhmann}. However, imperfect conductance has been measured in experiments performed on longer HgTe samples ~\cite{Konig2007,Roth,Konig2013,Gusev2011}, suggesting that TR-violating scatterers such as intrinsic nuclear spins or magnetic impurities may become important in such devices \cite{Lunde,Tarasenko,Koumoulis,Kurilovich,Russo,Hsu,Kimme,Zheng,Del Maestro}. Similar considerations are also relevant for 3D TI candidates such as $\text{Bi}_2\text{Se}_3$ \cite{Zhang2009,Xia2009} and $\text{Sb}_2\text{Te}_3$ \cite{Zhang2009}, which also include spinful nuclear isotopes and likely carry magnetic impurities as well \cite{Liu}. Both theoretical and experimental evidence that spinful nuclei lead to not only backscattering of helical modes but also dynamic nuclear polarization has also appeared recently~\cite{Del Maestro,Russo,Tian}. Despite this progress, it remains challenging to observe the precise role of nuclear spins or magnetic impurities in experiments because of the numerous other factors present in these devices.

In this paper, we investigate the impact of nuclear spins or magnetic impurities coupled to helical edge states in topological insulator nanorings. The introduction of a magnetic flux threading the ring provides an additional control knob to facilitate the study of the helical electron-impurity interaction. Quantum nanorings and disks have drawn a significant amount of attention over the past decade~\cite{Grujic,Ostahie,Rachel,Kikutake}, in part because they are ideal systems in which to study quantum interference phenomena such as the Aharonov-Bohm (AB) effect \cite{Aharonov,Borunda,Fomin,Viefers,Filusch} and other geometrical phase effects. Persistent currents in quantum rings due to the AB effect~\cite{Buttiker} were soon proposed after the AB effect itself and were confirmed experimentally~\cite{Webb}. Since then, persistent currents have been one of the most active fields of research in this context~\cite{Chandrasekhar,Lorke,Rabaud,Splettstoesser,Bayer,Fuhrer,Bleszynski,Bluhm,Ren}. Research on quantum rings is also expanding due to their various applications in spintronics. A few examples include spin entanglement control~\cite{Strom}, spin filtering~\cite{Popp}, spin beam splitting~\cite{Foldi2006} and spin current pumping~\cite{Citro}. Possible applications to quantum information processing have also been proposed ~\cite{Tatara}. Additional applications are possible in rings possessing a significant spin-orbit interaction \citep{Berche,Faizabadi,Nowak,Foldi2009}, and this has in part motivated recent investigations of TI quantum rings both theoretically~\cite{Jakovljevic} and experimentally~\cite{Autore}. The AB effect in such systems has been worked out theoretically and observed experimentally in both 2D and 3D TIs \cite{Konig2006,Bardarson,Gusev,Peng}. The bound-state spectrum of clean 2D TI quantum rings based on the BHZ model of HgTe quantum wells, in the presence of a magnetic field, has been calculated~\cite{Michetti}, but the effect of a magnetic impurity on this spectrum remains an open problem.
  
Here, we calculate the spectrum of helical edge states on a 2D TI ring coupled to a nuclear spin or magnetic impurity of arbitrary spin and with a magnetic flux threading the ring. Using a generalized time-reversal symmetry under which both the electronic and impurity spins are reversed, along with spin conservation, we derive a universal formula for the spectrum as a function of magnetic field that depends only on the amplitude of transmission through the impurity. Thus our results apply for any spatial profile of the electron-impurity interaction region. We show that the solution for an arbitrary-spin impurity can be built up using solutions for spin 1/2 and spin 1 impurities, which we obtain explicitly. We show that, in a certain energy regime, the spectrum becomes effectively independent of the magnetic flux for sufficiently strong impurity coupling, leading to sizable energy gaps. In addition, we calculate the entanglement entropy of the helical states as a function of magnetic field, finding that the electronic and impurity spins become maximally entangled near the spectral gaps.

The paper is structured as follows. In Section~\ref{sec_overview} we describe the TI ring-impurity model and discuss the symmetries present in this model and their consequences. We show that a generalized version of time-reversal symmetry allows us to decompose the model for an arbitrary-spin impurity into decoupled spin 1/2 and spin 1 sectors. Following this result, we solve the scattering problem for these two special cases in Sections~\ref{sec_half} and \ref{sec_integer}, respectively. In each case, we obtain a universal formula for the energy spectrum in terms of the transmission amplitude. In Section~\ref{sec_multi}, we discuss how our results generalize to the case of several impurities on the ring. In Section~\ref{sec_exp}, we calculate the spin current and entanglement entropy of the system. In Section~\ref{sec_gaps}, we study the dependence of the spectrum on the ring geometry and impurity couplings and obtain approximate formulas for energy bandwidths and gaps for a square impurity potential. Three appendices contain additional technical details pertaining to our derivations.


\section{Hamiltonian and symmetries} \label{sec_overview}

\subsection{TI ring in a magnetic field}

We take the non-interacting Hamiltonian on the ring to be the effective Hamiltonian of helical edge states in 2D TIs~\cite{Qi/Zhang}:
\begin{equation} 
\label{Hedge} H_0=v_0\hat p_y\sigma_z,
\end{equation}
where $v_0$ is the Fermi velocity, $\hat p_y$ is the (angular) momentum operator, and $\sigma_z$ acts on the spin subspace. The eigenstates of this Hamiltonian are spin-momentum locked plane waves: $\psi_+=e^{ip_yy}\ket{\uparrow}$, $\psi_-=e^{-ip_yy}\ket{\downarrow}$, where for a ring of circumference $d$, the momenta are quantized: $p_y=2\pi n/d$. We model the interaction between the electron spin and an impurity spin $I$ as
\begin{equation}
H_{S,I}= F(y)  \big[   A^z \sigma_zI_z + A^\perp (\sigma_-I_+ + \sigma_+I_-)  \big],\label{hyperfine}
\end{equation}
where we follow Ref.~\cite{Lunde} and use an interaction that has been averaged over impurity spin locations within the edge state. This interaction occurs over a finite region of width $w$ on the ring, with a specific spatial profile set by $F(y)$, and we allow for the longitudinal and transverse spin-spin coupling constants $A^z$ and $A^\perp$ to be different. This interaction breaks TR symmetry and provides a mechanism for backscattering that is assisted by electron-impurity spin flip-flops generated by the transverse terms in Eq.~\eqref{hyperfine}. 

\begin{figure}
\includegraphics[scale=1]{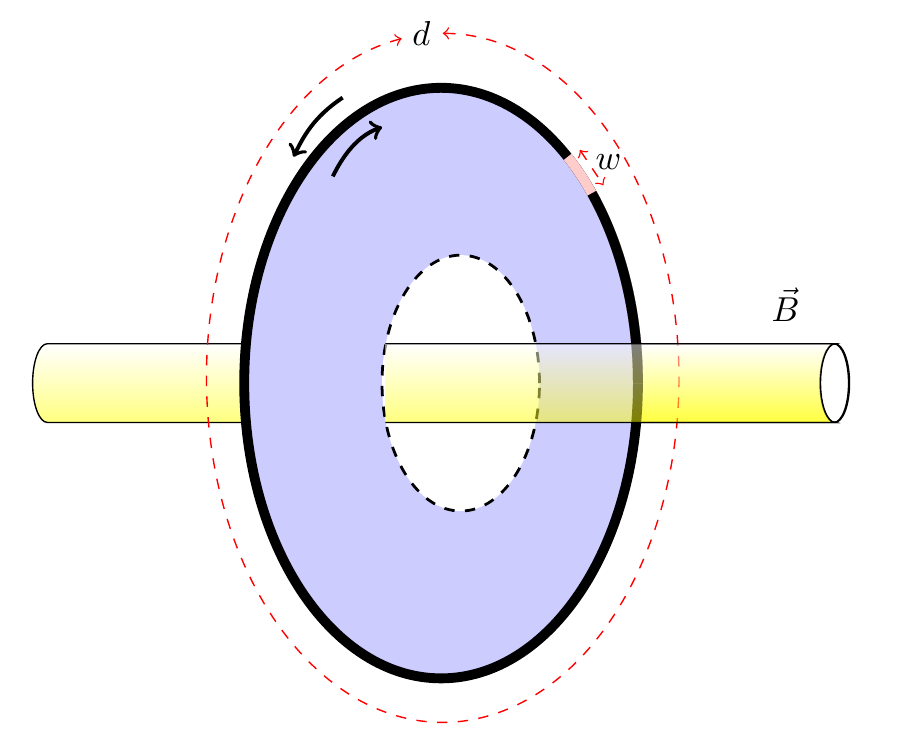}
\caption{(Color online)~A schematic illustration of a TI ring of circumference $d$ with a single impurity of width $w$. The edge states are indicated with arrows and the magnetic field associated with the AB effect is localized inside the ring.}
\label{fig-ring}
\end{figure}

In the presence of an applied magnetic flux $\Phi_B$ that threads the ring (see Fig.~\ref{fig-ring}), the momentum operator in Eq.~\eqref{Hedge} is shifted according to: $\hat p_y \to \hat p_y+p_B$, where $p_Bd=2\pi\Phi_B/\Phi_0$, where $\Phi_0$ is the magnetic flux quantum. This shift can be effectively undone by introducing an ansatz for the eigenstate wavefunction that includes a global phase:
\begin{equation} 
\label{psi_AB_y} \psi_{B}(y) = e^{-i p_B y} \psi(y).
\end{equation}
Here, $\psi(y)$ is a solution to the Hamiltonian without the vector potential, Eq.~\eqref{Hedge}, but now with a nontrivial boundary condition imposed on it as required to ensure single-valuedness of the wave function, $\psi_{B}(0)=\psi_{B}(d)$:
\begin{equation}
\psi(d)=e^{i p_B d} \psi(0)=e^{2\pi i \Phi_B/\Phi_0} \psi(0).\label{singlevalued}
\end{equation}
The first equality resembles a Bloch constraint in one dimension, with $d$ interpreted as the lattice spacing and $p_B$ as the crystal momentum. Thus, we can treat the impurity scattering problem with nonzero magnetic flux as though we are solving a Kronig-Penney model with Hamiltonian $H=H_0+H_{S,I}$ {\it without} magnetic flux. The magnetic flux dependence is restored by replacing the crystal momentum by $p_B$. This observation reflects the general connection between the AB problem on a ring with a non-uniform potential and the Kronig-Penney model~\cite{Romanovsky,Buttiker,Ghosh}. We may thus think of the energy spectrum dependence on $p_B$ or $\Phi_B$ as an effective band structure.

\subsection{Spin conservation and generalized time-reversal symmetry}

Although TR symmetry is broken by the coupling to the impurity, Eq.~\eqref{hyperfine}, this interaction does preserve a generalized time-reversal (GTR) symmetry that flips both the electronic and impurity spins. By exploiting this symmetry along with conservation of total spin, we can achieve a better understanding of the boundary matching problem that we must solve in order to obtain the energy spectrum. In fact, the consequences of these two symmetries together lead to a universal result for the energy spectrum in terms of only one variable, which we take to be the amplitude of transmission through the impurity. 

\begin{figure}
\includegraphics[scale=1]{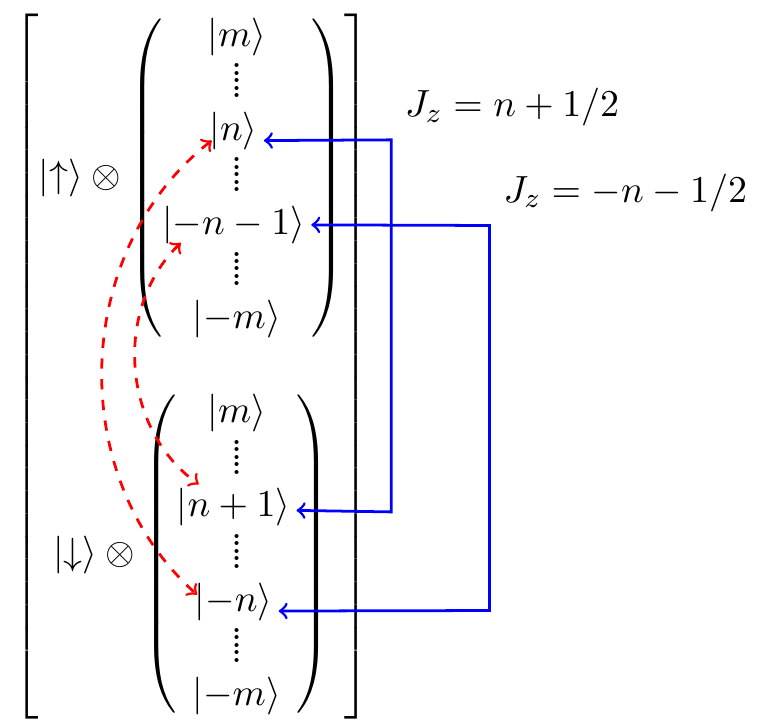}
\caption{(Color online)~Decomposition of the wavefunction into its sectors and blocks (here we suppress the spatial dependence). $\ket{\uparrow}$ and $\ket{\downarrow}$ are electron spin states, and $\ket{n}$ labels impurity spin states. Two components with the same total spin $J_z$ that form one sector are connected with blue arrows. Red arrows show the components that mix under GTR. We see that the two sectors $J_z=n+1/2$ and $J_z=-n-1/2$ mix and form the block labeled by $J_z=|n+1/2|$.}
\label{fig-blocks}
\end{figure}

The total Hamiltonian, $H=H_0+H_{S,I}$, commutes with the total spin operator $J_z=S_z+I_z$. Thus, the total wavefunction describing both the electron and impurity breaks into sectors labeled by $J_z$.  Two of these sectors (the ones corresponding to maximum and minimum $J_z$) are one-dimensional, while the rest are two-dimensional, as is illustrated in Fig.~\ref{fig-blocks}. 

To understand the consequence of GTR, first note that this operation maps the electron-impurity spin state $\ket{\uparrow}\ket{n}$ to $\ket{\downarrow}\ket{-n}$. The former state has $J_z=n+1/2$, while the latter has $J_z=-n-1/2$, so we see that GTR mixes different total spin sectors. Importantly, it mixes {\it only} these two sectors, so that the Hilbert space breaks up into ``blocks'', where each block consists of two sectors of opposite $J_z$ and thus can be labeled by $|J_z|$ (Fig.~\ref{fig-blocks}) \footnote{This whole concept is similar to the BHZ model of HgTe quantum wells in which the two blocks of the Hamiltonian are related through time-reversal~\cite{BHZ}. Furthermore, similar spin conservation arguments arise when we include spin-orbit interactions in this model~\cite{Rothe,Michetti,Lunde}.}. Of course, when $J_z=0$ there is only a single sector in the block; this type of block only occurs for half-integer-spin impurities. In terms of scattering eigenstates, GTR relates an eigenstate incoming from one side of the impurity to an eigenstate incoming from the opposite side, and it allows us to solve for these eigenstates by separately solving the matching problem in each block. In the next two sections, we exploit this fact to derive general relations between the reflection and transmission amplitudes for a single impurity. We then use these relations to obtain a universal formula for the energy spectrum; we find that the same formula arises in every block regardless of the value of $|J_z|$. Thus, the entire spectrum for an arbitrary-spin impurity can be obtained from this formula after the transmission amplitudes in each sector are calculated for a given interaction region profile $F(y)$.

\section{Spin-1/2 impurity} \label{sec_half}

In this section, we solve the scattering problem for a single impurity with spin $m=1/2$. We begin with ansatz eigenstate wavefunctions on the left (L) and right (R) sides of the interaction region:
\begin{eqnarray} 
\label{Hsol2} \psi_{L}(y)&=&e^{i p y} \begin{pmatrix}\alpha\\\beta\\0\\0\end{pmatrix} +e^{-i p y} \begin{pmatrix}0\\0\\\alpha'\\\beta'\end{pmatrix},\nonumber\\  \psi_{R}(y)&=&e^{i p y} \begin{pmatrix}\alpha''\\\beta''\\0\\0\end{pmatrix} +e^{-i p y} \begin{pmatrix}0\\0\\\alpha'''\\\beta'''\end{pmatrix}.
  \end{eqnarray}
Here, the basis states for the spinors are (from top to bottom) $\ket{\uparrow,1/2}$, $\ket{\uparrow,-1/2}$, $\ket{\downarrow,1/2}$, $\ket{\downarrow,-1/2}$, where the arrows denote the electron spin, and $\pm1/2$ refers to the impurity spin. The coefficients $\alpha$ and $\beta$ specify the ``initial'' impurity spin state for an eigenstate incoming from the left, while $\alpha'''$ and $\beta'''$ give the initial impurity state for an eigenstate incoming from the right. The remaining coefficients $\alpha'$, $\beta'$, $\alpha''$, $\beta''$ correspond to reflection or transmission coefficients, depending on the direction from which the incident wave originates.

\begin{figure}
\includegraphics[scale=1]{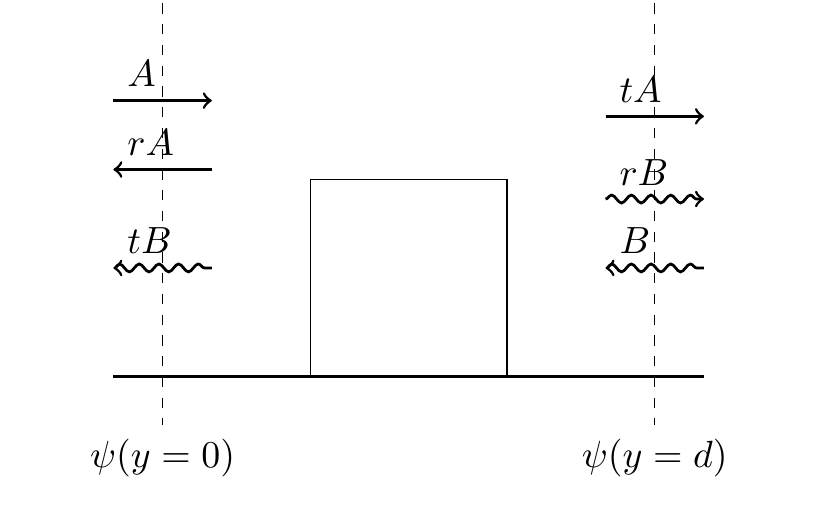}
\caption{Scattering off a single impurity (depicted here as a square barrier, which is assumed to carry nonzero spin $I$). Incident waves from the left (with coefficient $A$) and right (with coefficient $B$) and their corresponding transmitted and reflected waves are shown. We assume the impurity is in an eigenstate of $I_z$ with eigenvalue $|n|<I$. Each arrow corresponds to one component of the wavefunction (on left or right) with a particular spin state. Combinations of these waves form the wavefunctions in Eq.~(\ref{rtwf1/2}).}
\label{fig-Incidents}
\end{figure}

The standard approach to solving this type of scattering problem is to also solve for the wavefunction inside the interaction region and to then enforce continuity of the wavefunction at the boundaries of this region. Imposing the additional constraint, Eq.~\eqref{singlevalued}, generally gives rise to a relation between the energy and the magnetic flux, leading to the effective band structure $E(\Phi_B)$.

Here, we describe a more economical approach to obtaining the energy spectrum that yields a universal formula for arbitrary-spin impurities with the help of GTR symmetry. We first demonstrate this for a spin 1/2 impurity and postpone discussion of larger spins to the next section. First note that the wavefunction spinor components with $J_z=\pm1$ do not mix with each other or with any other components under the electron-impurity interaction, Eq.~\eqref{hyperfine}, and simply acquire phase factors as a consequence of the impurity. These phases have no bearing on the energy spectrum, and thus the $|J_z|=1$ block can be safely neglected. (The impact of GTR on these phases is discussed in Appendix~\ref{1/2_t_r_appendix}.) We therefore focus only on the $J_z=0$ block. If our impurity scattering problem were formulated in an infinite 1D channel instead of on a finite ring, then every left-incoming eigenstate would be degenerate with a right-incoming state, and any  superposition of these would also be an eigenstate. On the left and right side of the impurity barrier (see Fig.~\ref{fig-Incidents}), we could then write 
\begin{equation} 
\label{rtwf1/2}  \psi(0)=\begin{bmatrix} A \\ r_\rightarrow A+t_\leftarrow B \end{bmatrix},\quad  \psi(d)=\begin{bmatrix} t_\rightarrow A+r_\leftarrow B \\ B \end{bmatrix}, \end{equation}
where $A$ and $B$ are the coefficients of the left-incoming and right-incoming states in the superposition, and we only keep the $J_z=0$ wavefunction spinor components $\ket{\uparrow,-1/2}$ and $\ket{\downarrow,1/2}$. The subscript arrows on the reflection and transmission amplitudes indicate the direction of the corresponding incoming wave. If we now return to the ring geometry by identifying $y=0$ and $y=d$ and imposing the single-valuedness constraint, Eq.~\eqref{singlevalued}, then we find that only one of these superpositions is a valid eigenstate, and the magnetic flux is determined by the scattering amplitudes:
\begin{widetext}
\begin{eqnarray}
B/A&=&\frac{1-r_{\leftarrow}r_{\rightarrow}-t_{\leftarrow}t_{\rightarrow}\pm\sqrt{(1-r_{\leftarrow}r_{\rightarrow}+t_{\leftarrow}t_{\rightarrow})^2-4t_{\leftarrow}t_{\rightarrow}}}{2r_{\leftarrow}t_{\leftarrow}},\\
e^{2\pi i \Phi_B/\Phi_0}&=&\frac{1-r_{\leftarrow}r_{\rightarrow}+t_{\leftarrow}t_{\rightarrow}\pm\sqrt{(1-r_{\leftarrow}r_{\rightarrow}+t_{\leftarrow}t_{\rightarrow})^2-4t_{\leftarrow}t_{\rightarrow}}}{2t_{\leftarrow}}.\label{bandstructure1}
\end{eqnarray}
\end{widetext}
The two solutions distinguished by the sign in front of the square root correspond to currents circulating in opposite directions around the ring, as we discuss further in Sec.~\ref{sec_exp}. These two solutions are degenerate and are related to each other by GTR symmetry. It is important to note that Eq.~\eqref{bandstructure1} holds for {\it any} barrier shape $F(y)$; the only assumption we have made is that the barrier vanishes at $y=0$ and $y=d$. Once the scattering amplitudes are obtained for a given barrier shape, Eq.~\eqref{bandstructure1} can be used to obtain the corresponding energy spectrum.

\begin{figure}
    \centering	
   \includegraphics[scale=.95]{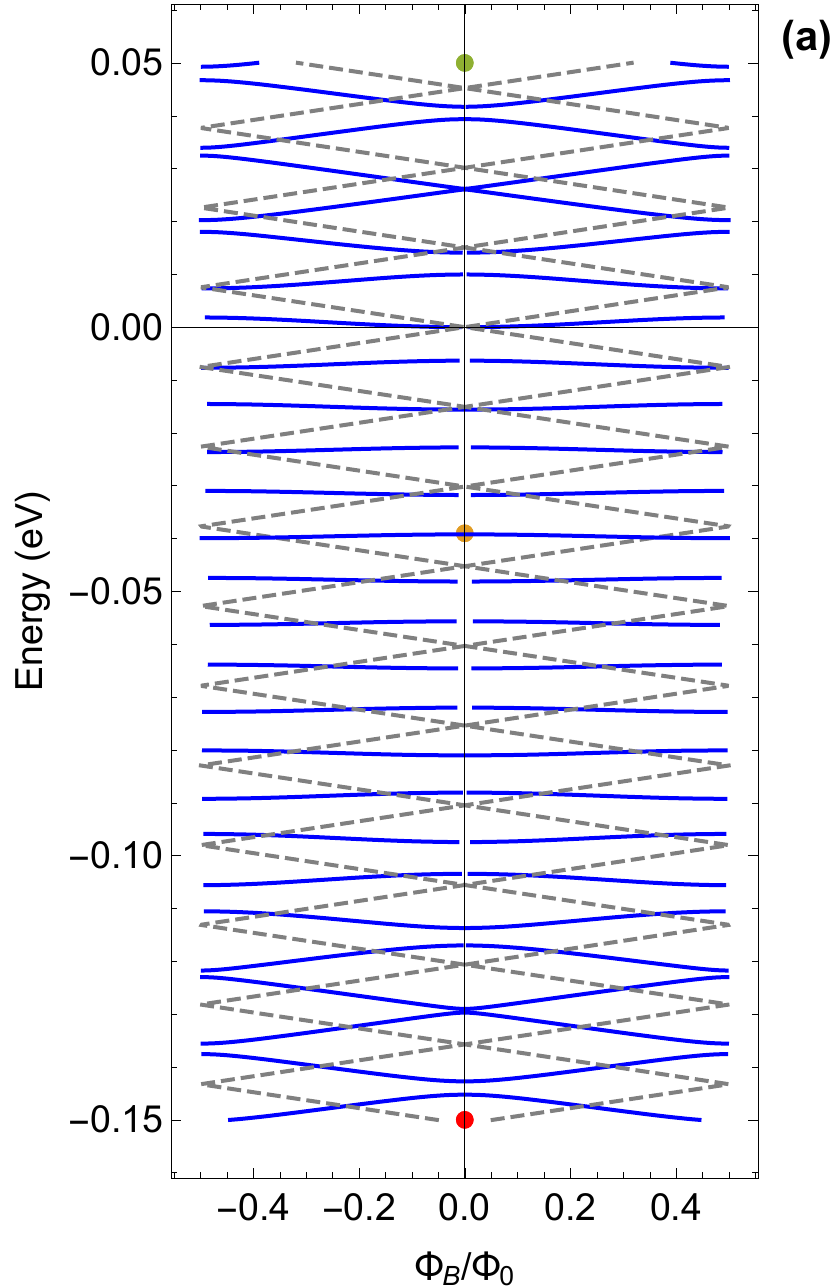}
    \includegraphics[scale=.95]{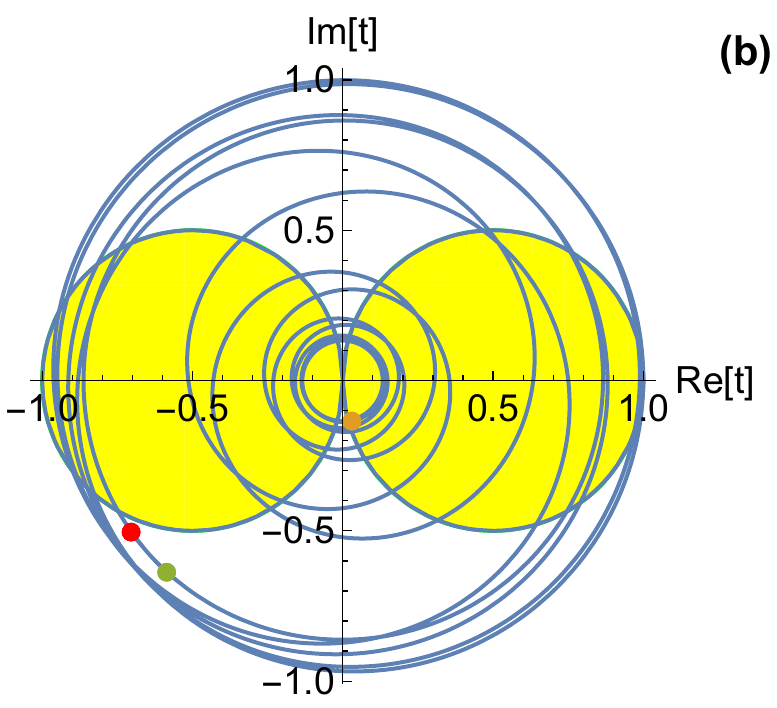}
\caption{(Color online)~(a) Energy spectrum (solid blue lines) as a function of magnetic flux for a spin 1/2 impurity with square potential. The dashed lines indicate the spectrum without the impurity. (b) Parametric plot of the transmission amplitude as a function of energy. Yellow regions indicate values where the condition $\cos^2\phi_t<|t|^2$ does not hold and thus correspond to gaps in the spectrum. Colored dots map certain energies of the energy spectrum to points on the parametric plot. The parameters are $d=1000$ \AA, $w=130$ \AA, $A^z=A^\perp=0.05~\text{eV}$ \cite{Liu}, $v_0=2.4~\text{eV \AA}$ \cite{Lunde}.}
\label{fig-spin-half-band}
\end{figure}

In the absence of GTR or any other symmetry, the reflection and transmission amplitudes $r_{\rightarrow}$, $r_{\leftarrow}$, $t_{\rightarrow}$, $t_{\leftarrow}$ would all be independent of each other (aside from the normalization condition). However, as explained in detail in Appendix~\ref{1/2_t_r_appendix}, GTR symmetry imposes two relations among these amplitudes:
\begin{equation}
t_\leftarrow=t_\rightarrow\equiv t\equiv|t|e^{i \phi_t},\quad
\label{rts1/2}  r_\rightarrow r_\leftarrow= (|t|^2-1)e^{2i\phi_t}.
\end{equation}
These relations dramatically simplify Eq.~\eqref{bandstructure1} and allow us to express the magnetic flux in terms of only the transmission amplitude:
\begin{equation} 
\label{lambda1/2}e^{2\pi i \Phi_B/\Phi_0}= \cos\phi_t/|t| \pm \sqrt{(\cos\phi_t/|t|)^2 -1 }.
\end{equation}
Since the left-hand-side is a pure phase, the right-hand-side must also be a pure phase in order for a solution to exist. This then leads to the following condition for a state to exist at a given energy:
\begin{equation} 
\label{cond1/2}\cos^2\phi_t<|t|^2.
\end{equation}
Energy ranges where $t(E)$ violates this condition correspond to gaps in the spectrum. In ranges where states exist, the two different branches of the square root in Eq.~\eqref{lambda1/2} simply correspond to the fact that the energy is independent of the sign of the magnetic flux: $E(\Phi_B)=E(-\Phi_B)$. For reasons that will become clear in the next section, we refer to this feature as a 2-fold ``flux degeneracy'', i.e., the number of distinct values of the flux $\Phi_B$ that give rise to a state of a given energy $E$. Fig.~\ref{fig-spin-half-band}(a) shows an example band structure obtained from this formula for the case of a square barrier using typical experimental parameters. The scattering amplitudes for the square barrier are derived in Appendix~\ref{appendix_general_rt}. One salient feature of the spectrum is that band edges always occur at half-integer multiples of the flux quantum. This generally holds for any spin 1/2 impurity regardless of couplings or potential shape and follows directly from the band edge condition $\cos\phi_t=\pm|t|$ and Eq.~\eqref{lambda1/2}. We will see in the next section that band edges can occur at other values of $\Phi_B$ for higher-spin impurities. The most striking consequence of the impurity is the occurrence of nearly flat bands in the vicinity of $E=-A^z=-0.05$ eV, with gaps of size 8 meV between them. We explain the origin of these bands and study their dependence on the impurity couplings and ring geometry in Sec.~\ref{sec_gaps}. We further analyze the spectrum for a spin 1/2 impurity quantitatively for a range of realistic device parameters in the same section.

The constraint in Eq.~(\ref{cond1/2}) can be visualized in terms of the complex $t$ plane (Fig.~\ref{fig-spin-half-band}(b)). Scanning through values of the energy corresponds to tracing out a curve in this plane, and whenever the curve enters one of the yellow regions, which indicate values of $t$ that violate Eq.~\eqref{cond1/2}, a gap occurs in the spectrum. Large loops give rise to dispersive bands, while the flat bands correspond to loops concentrated close to the origin. If the parametric curve tangentially touches the yellow region at $\hbox{Re}[t]=\pm1$, then a band touching point appears in the spectrum at $\Phi_B=0$ or $\pm1/2$. Such points can only occur at values of the energy for which $|t|=1$, i.e., for which the impurity is effectively transparent. We show in Sec.~\ref{sec_gaps} that a discrete set of energies satisfy this condition in the case of a square impurity barrier.

\section{Arbitrary-spin impurity} \label{sec_integer}

In this section, we solve the scattering problem for an arbitrary-spin impurity by following an approach that is similar to what we used for a spin 1/2 impurity in the previous section. Since GTR symmetry allows us to solve the problem in each block of the Hilbert space separately, we only need to solve the case of a spin 1 impurity to obtain the solution in the general case. The spin 1 case has only one block consisting of two GTR-coupled sectors. Impurities with larger integer spins will break into a series of blocks, all with the same structure as the spin 1 case, allowing us to solve these cases in terms of multiple copies of the spin 1 impurity solution. Moreover, half-integer spins will also reduce to one spin 1/2 block (with $J_z=0$) and several spin 1 blocks (with $J_z \neq 0$). As a result, we can solve the problem for an arbitrary-spin impurity by combining the solutions for the spin 1/2 and spin 1 cases.

Out of the six components of the electron-impurity wavefunction for a spin 1 impurity, the two with maximal $|J_z|$ (i.e., $J_z=\pm3/2$) are again decoupled from each other and from all other components, while the remaining four components form the block with $|J_z|=1/2$. From now on we distinguish all variables of this block with $\pm$ signs to indicate the sector to which it belongs according to the sign of $J_z$. As in the previous section, we consider left-incoming waves in each sector with coefficients $A^\pm$ superposed with right-incoming waves with coefficients $B^\pm$. The analog of Eq.~(\ref{rtwf1/2}) becomes
\begin{equation} 
\label{rtwf1}  \psi(0)=\begin{bmatrix}A^+ \\ A^-\\ r_{\rightarrow}^+ A^+ +t_{\leftarrow}^+B^+ \\ r_{\rightarrow}^- A^-+t_{\leftarrow}^- B^- \end{bmatrix}, \quad \psi(d)= \begin{bmatrix} t_{\rightarrow}^+A^+ +r_{\leftarrow}^+B^+\\ t_{\rightarrow}^-A^-+r_{\leftarrow}^-B^-\\B^+\\B^- \end{bmatrix},
\end{equation}
where the basis states are now $\ket{\uparrow,0}$, $\ket{\uparrow,-1}$, $\ket{\downarrow,1}$, $\ket{\downarrow,0}$. Since $\psi(0)$ and $\psi(d)$ are each essentially just two copies of the analogous expressions in the spin 1/2 case, Eq.~\eqref{rtwf1/2}, when we impose the single-valuedness condition, Eq.~\eqref{singlevalued}, we obtain two copies of the band structure equation, Eq.~\eqref{bandstructure1}, one for each sector labeled by $\pm$. 

\begin{figure}
    \centering	
 \hspace*{0cm} \includegraphics[width=0.9\columnwidth]{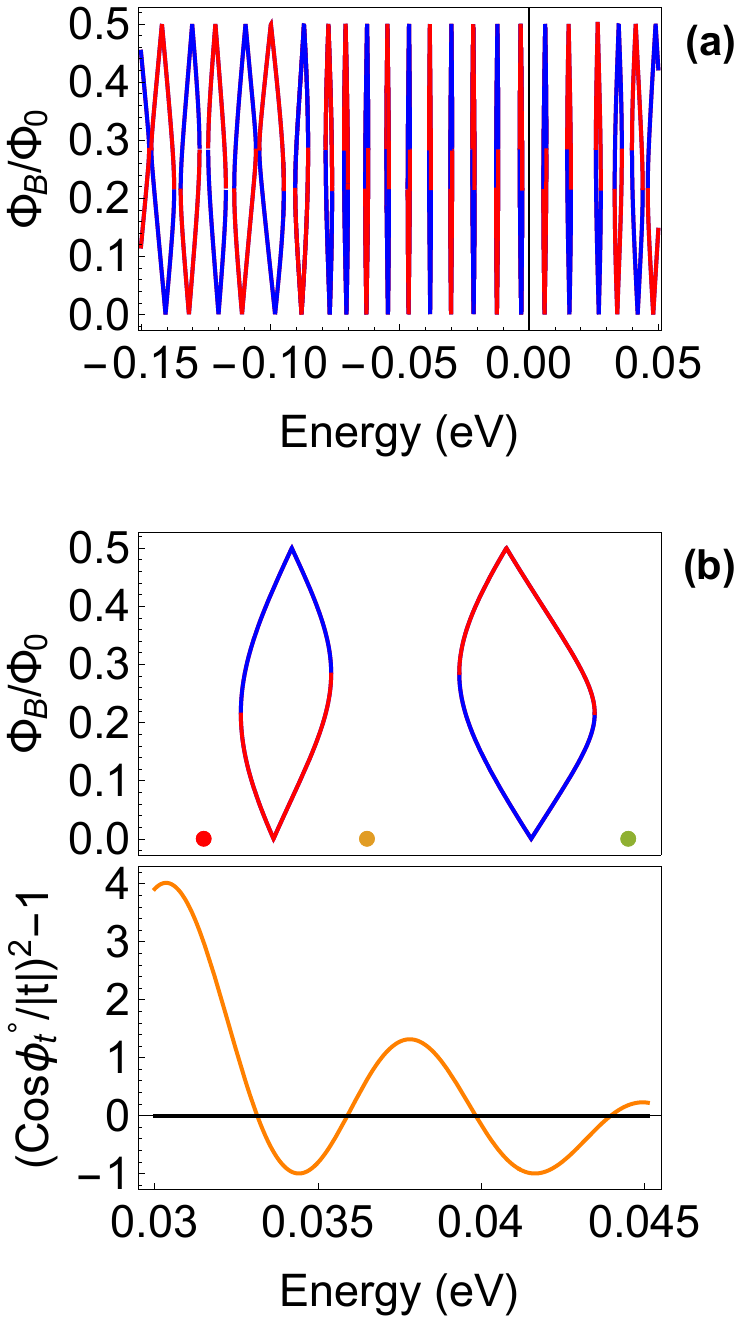}\vspace{-0.1cm}
      \hspace*{0cm}  \includegraphics[width=0.9\columnwidth]{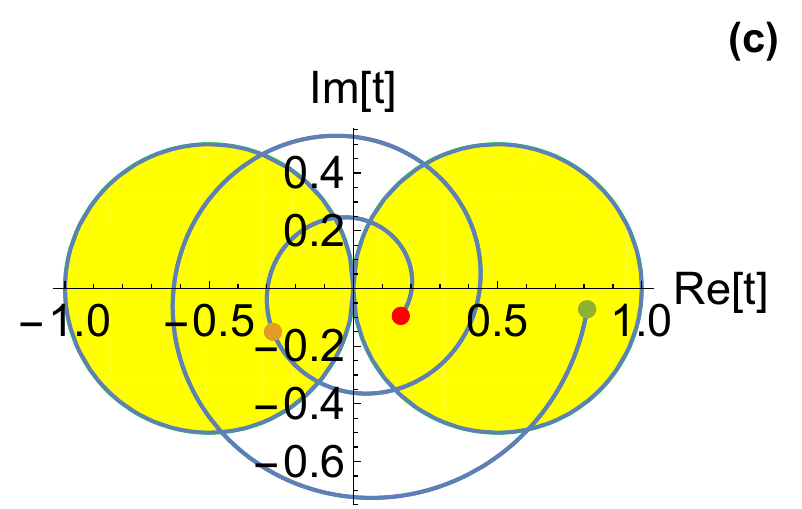}
\caption{(Color online)~Spin 1 impurity. (a) Energy spectrum as a function of magnetic flux. Each of the two total spin sectors yields a different state (distinguished by red and blue) at each energy. (b) Zoom-in of the spectrum shown in (a) (upper panel) and the condition for states to exist (lower panel). States occur at energies where the transmission amplitudes satisfy $|\cos\phi_t^\circ|<|t|$. (c) Parametric plot of transmission amplitude $t=|t|e^{i\phi_t^\circ}$ for the energy range shown in (b). The colored dots indicate the corresponding energies shown in (b). The parameters are $d=1000$ \AA, $w=130$ \AA, $A^z=A^\perp=0.05~\text{eV}$ \cite{Liu}, $v_0=2.4~\text{eV \AA}$ \cite{Lunde}.}
\label{fig-spin-one-band}
\end{figure}

We may again invoke GTR symmetry to simplify these expressions using relations between the scattering amplitudes. However, since GTR now couples two distinct sectors, this process is different from the spin 1/2 case, for which there was only a single sector. The details are given in Appendix~\ref{1_t_r_appendix}. The resulting relations among the amplitudes are as follows:
\begin{eqnarray}\label{rtlr1}
 &&t_\rightarrow^+= t_\leftarrow^-, \quad  t_\leftarrow^+= t_\rightarrow^-, \quad
r_\rightarrow^+ = r_\rightarrow^- , \quad r_\leftarrow^+ = r_\leftarrow^-,\nonumber\\
&&|t^\pm_\rightarrow|=|t^\pm_\leftarrow|\equiv|t|,\quad |r^\pm_\rightarrow|=|r^\pm_\leftarrow|\equiv|r|,\quad |r|^2+|t|^2=1,\nonumber\\&&
\end{eqnarray} 
and additionally we have 
\begin{equation} \label{rtwf1}
\phi_{t^\pm}^\circ = \phi_{r^\pm}^\circ +\pi/2, 
\end{equation}
where we have defined,
\begin{equation}\label{avgphase}
\phi_x^\circ \equiv (\phi_{x\rightarrow} + \phi_{x\leftarrow})/2, \quad \text{for} \quad x=r^\pm,t^\pm,
\end{equation}
where $\phi_{x\rightarrow}$, $\phi_{x\leftarrow}$ are the phases of the corresponding scattering amplitudes. It follows from Eq.~\eqref{rtlr1} that $\phi_{t^+}^\circ=\phi_{t^-}^\circ$, which allows us to drop the sector labels $\pm$ in these quantities: $\phi_{t^+}^\circ=\phi_{t^-}^\circ \equiv \phi_{t}^\circ $. These relations allow us to simplify Eq.~\eqref{bandstructure1} down to the result
\begin{equation} \label{lambda1} 
e^{2\pi i \tfrac{\Phi_B}{\Phi_0}} =e^{i (\phi_{t^\pm_\rightarrow}- \phi_{t^\pm_\leftarrow})/2}\left[\cos\phi_t^\circ/|t| \pm \sqrt{(\cos\phi_t^\circ/|t|)^2-1}\right].
\end{equation} 
The overall phase factor in this expression depends on the sector as can be seen from Eq.~\eqref{rtlr1}; the two phases in fact differ only by a sign: $\phi_{t^+_\rightarrow}- \phi_{t^+_\leftarrow}=-(\phi_{t^-_\rightarrow}- \phi_{t^-_\leftarrow})$. An important consequence of the overall phase is that there is now a 4-fold flux degeneracy instead of a 2-fold degeneracy like we have for a spin 1/2 impurity. This 4-fold degeneracy comes from the two sectors and the two branches of the square root in Eq.~\eqref{lambda1}. Because the overall phase differs by only a sign between the two sectors, it remains true that $E(-\Phi_B)=E(\Phi_B)$, or in other words the spectrum remains symmetric about $\Phi_B=0$. An example spectrum for a spin 1 impurity is shown in Fig.~\ref{fig-spin-one-band}(a), where the additional degeneracy is evident. Also notice that, unlike in the spin 1/2 impurity case, the band edges can occur at arbitrary values of the flux; this is due to the extra phase $e^{i (\phi_{t^\pm_\rightarrow}- \phi_{t^\pm_\leftarrow})/2}$ appearing in Eq.~\eqref{lambda1}. We also note that flat bands are again apparent in the region near $E=-A^z$.

Also notice the similarity of Eq.~\eqref{lambda1} to Eq.~\eqref{lambda1/2}. Aside from the overall phase factor, the only other difference is that $\phi_t$ has been replaced by the average phase $\phi_t^\circ$. Eq.~\eqref{lambda1/2} can be understood as a special case of Eq.~\eqref{lambda1} where the self-duality of the $J_z=0$ sector GTR symmetry enforces $t_\rightarrow=t_\leftarrow$, so that the overall phase factor in Eq.~\eqref{lambda1} vanishes, and $\phi_t^\circ$ reduces to $\phi_t$. 

All of our analysis here was based only on the fact that two sectors are mixed by GTR symmetry (except when $J_z=0$ where there is only one self-symmetric sector). This is true for any value of $J_z$ so that in fact Eq.~(\ref{lambda1}) holds regardless of the spin of the impurity. Thus we conclude that the flux degeneracy for an impurity of spin $m$ is equal to twice the number of distinct nontrivial sectors (i.e., those with $|J_z|\ne m+1/2$), since each such sector contributes two solutions corresponding to the two branches of the square root in Eq.~\eqref{lambda1}. Hence the flux degeneracy is $2(2m+1)-2=4m$.  

Note that even though the spectrum formula, Eq.~\eqref{lambda1}, holds regardless of the spin of the impurity, the resulting energy spectra still depend sensitively on the spin and potential of the impurity since these details strongly affect the transmission amplitudes (i.e., the $t$'s) that enter this formula. These amplitudes are determined by diagonalizing of the Hamiltonian, which depends on the impurity couplings and potential. As an example, consider a spin 3/2 impurity. In this case, there are three sectors ($J_z=0$ and $J_z=\pm1$), and each produces a unique spectrum formula condition like Eq.~\eqref{lambda1}. To obtain the spectrum, it is necessary to diagonalize the total Hamiltonian in each of the two blocks ($J_z=0$ and $|J_z|=1$), extract the transmission amplitudes, and plug them into Eq.~(\ref{lambda1}).


\section{Multiple impurities} \label{sec_multi}

\begin{figure}
\includegraphics[width=0.9\columnwidth]{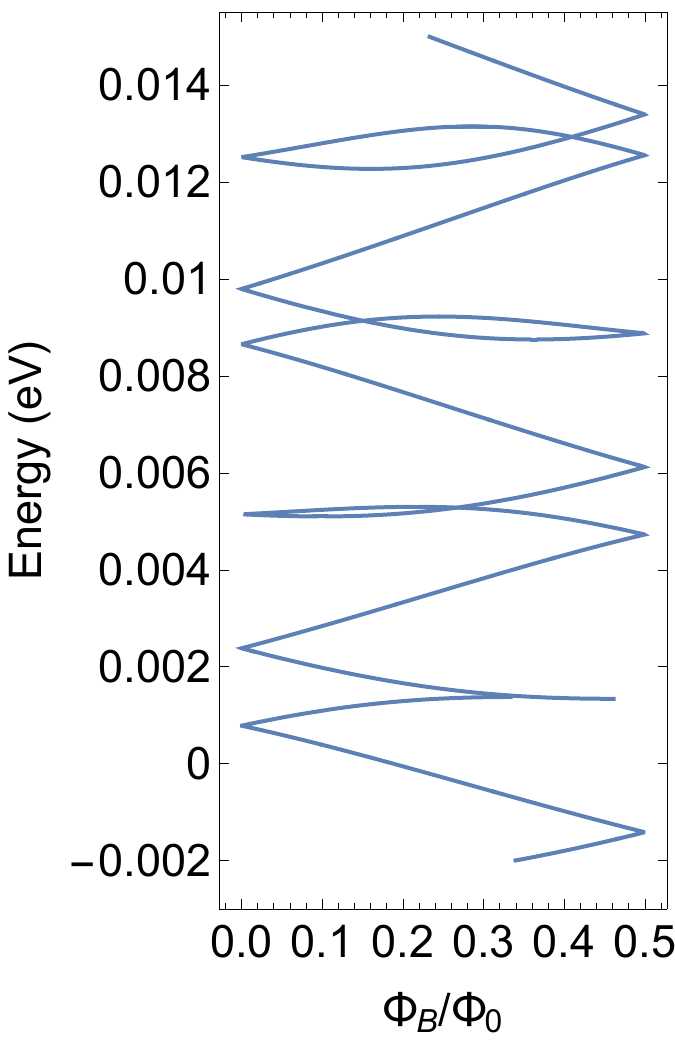}
\caption{(Color online)~Energy spectrum for two spin 1/2 impurities as a function of magnetic flux for parameters $d=1000$ \AA, $A^z = 0.05~\text{eV}$, $A^\perp=0.05~\text{eV}$, $w=130$ \AA, and $v_0=2.4~\text{eV \AA}$. }
\label{fig-multi}
\end{figure}

To treat the case of $N$ impurities on the ring, we can proceed in the same way as for a single impurity. In particular, we can begin by writing down ansatz wavefunctions at $y=0$ and $y=d$, each of which now contains $2(2I+1)^N$ spinor components assuming each impurity has the same total spin $I$. The Hilbert space again divides into sectors labeled by total spin $J_z$, where now the dimensions of the sectors ${\cal D}_{J_z}$ depend on $J_z$. In each sector, we take the ansatz wavefunctions to be superpositions of left-incoming and right-incoming states, and we express these wavefunctions in terms of reflection and transmission coefficients as in Eq.~\eqref{rtwf1}. We then apply the single-valuedness condition, Eq.~\eqref{singlevalued}, separately in each sector. Doing so will yield a polynomial in $e^{2\pi i \Phi_B/\Phi_0}$ for each sector, where the order of this polynomial is the dimension of that sector, ${\cal D}_{J_z}$. As an example, consider $N$ spin 1/2 impurities, for which the dimension of each sector is the binomial coefficient
\begin{equation} \label{comb}
 {\cal D}_{J_z}=\binom{N+1}{J_z+\frac{N+1}{2}}.
\end{equation}
For instance for $N=2$, the number of states with $J_z=+1/2$ is $\binom{3}{2}=3,$ which means that we have to solve a cubic equation in order to find the spectrum (recall that for one impurity, the resulting polynomial was quadratic and led to Eq.~\eqref{lambda1/2}). As we add more impurities, the order of this polynomial grows exponentially, and it quickly becomes necessary to solve for the spectrum numerically. An example of a spectrum for two spin 1/2 impurities (both with square potentials of equal size $w$ and with equal couplings) computed in this way is shown in Fig.~\ref{fig-multi}. In this case, there is up to a six-fold flux degeneracy depending on the energy. Unlike in the case of a single impurity, here the spectrum no longer exhibits flat bands in the vicinity of $E=-A^z$.


\section{Spin current and entanglement entropy} \label{sec_exp}

\subsection{Probability and spin currents}

In this section, we show that the symmetries of the TI ring-impurity system also lead to a universal formula for the ratio of spin and probability currents. The probability current is given by
\begin{equation} \label{jpd} 
j_p=v_0\psi(y)^\dagger \sigma_z \psi(y),  
\end{equation}
while the spin current is
\begin{equation} \label{jsd} 
j_s=(v_0/2)\psi(y)^\dagger \psi(y).  
\end{equation}
In the case of a spin 1/2 impurity, it is straightforward to find the ratio of these two quantities using the wavefunction ansatzes in Eq.~\eqref{rtwf1/2} in conjunction with the single-valuedness condition, Eq.~\eqref{singlevalued}:
\begin{equation} \label{jsp1/2}  
j_s=j_p\frac{1}{2} \frac{|1 -e^{2\pi i \Phi_B/\Phi_0} t|^2+|r|^2}{|1 -e^{2\pi i \Phi_B/\Phi_0} t|^2-|r|^2}.   
\end{equation}
Note that this expression is independent of the coefficients $A$ and $B$ that we introduced in Eq.~(\ref{rtwf1/2}). In this expression, $e^{2\pi i \Phi_B/\Phi_0}$ also depends on $r$ and $t$, and therefore the right hand side of the equation only depends on energy. For a spin 1 impurity (or more generally for one block of a larger-spin impurity), a similar expression can be derived (see Appendix~\ref{1_t_r_appendix}):
\begin{equation} \label{jsp1} 
j_s =j_p \frac{1}{2} \frac{|1-e^{2\pi i \Phi_B/\Phi_0} t_\leftarrow^+|^2+|r|^2}{|1-e^{2\pi i \Phi_B/\Phi_0} t_\leftarrow^+|^2-|r|^2}.
\end{equation}
Although it appears that the right-hand side depends on the sector (i.e., on total spin $J_z$), this is in fact not the case, as is shown in Appendix~\ref{1_t_r_appendix}. Hence, the currents in both sectors of the block are identical. A detailed discussion of spin current pumping in TIs can be found in Ref.~\cite{Citro2011}. We are assuming that the magnetic field is fully localized inside the ring so that there is no Zeeman interaction for either the electronic or impurity spins. It is readily apparent from Eqs.~\eqref{jsp1/2} and \eqref{jsp1} that the ratio of the currents can be controlled directly by adjusting the magnetic flux.


\subsection{Entanglement entropy of nuclear spin and electron}

\begin{figure*}
\includegraphics[scale=1]{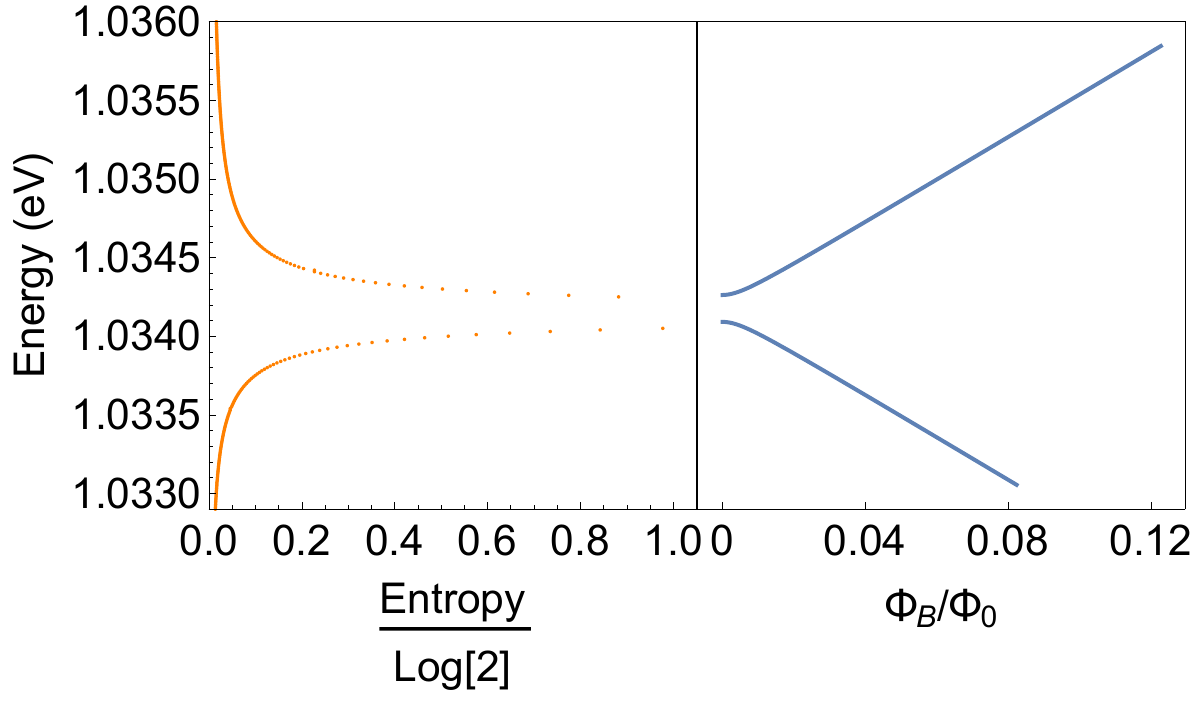}
\caption{(Color online)~Comparison of the energy spectrum (right) with the electron-impurity entanglement entropy (left) for a spin 1/2 impurity with square potential. Near the spectral gaps, the entanglement reaches its maximum possible value of $\log2$. The parameters are $d=1000$ \AA, $w=130$ \AA, $A^z=A^\perp=0.05~\text{eV}$ \cite{Liu}, $v_0=2.4~\text{eV}$~\AA \cite{Lunde}.}
\label{fig-entropy}
\end{figure*}

Next, we analyze the entanglement between the electron and impurity as a function of magnetic flux. We can define a position-independent entanglement entropy between electron and impurity spins in the following way. We begin by writing the TI ring-impurity eigenstates as
\begin{equation} \label{Hsol_sum_2}
\ket{\Psi} =\int dy \sum_{ij} \psi_{ij}(y) \ket{y,i,j}, 
\end{equation}
where the index $i$ denotes the electron spin state and $j$ the nuclear spin state. The full density matrix is given by $\rho_{I,e}= \ket{\Psi} \bra{\Psi}$. After tracing out the electronic spin, $\rho_{I}= \sum_i\bra{i} \rho_{I,e} \ket{i},$ this quantity will only depend on the impurity spin and the electron position, which we integrate out:
\begin{equation} \label{Hsol_sum_3}
\rho_{I}=\sum_{i}\int dy\sum_{j,j'} \psi_{ij}(y)\bar{\psi}_{ij'}(y)\ket{j}\bra{j'}. 
\end{equation}
Since the spin states $\ket{\uparrow,1/2}$ and $\ket{\downarrow,-1/2}$ do not mix with other states, we drop them and focus on the spin states in the $J_z=0$ sector:
\begin{equation} \label{rho_S} 
\rho_{I}=\begin{pmatrix}\int_{0}^{d} dy |\psi_{\uparrow,-1/2}|^2 & 0  \\0  & \int_{0}^{d} dy |\psi_{\downarrow,1/2}|^2 \end{pmatrix}. 
\end{equation}
To simplify this result further, notice that the probability current can be written as,
\begin{equation}
(1/v_0)j_p=|\psi_{\uparrow,-1/2}|^2 - |\psi_{\downarrow,1/2}|^2,
\end{equation}
which should be constant over the entire ring. This in turn implies
\begin{equation}
(d/v_0)j_p=\int_{0}^{d} dy|\psi_{\uparrow,-1/2}|^2 -\int_{0}^{d} dy |\psi_{\downarrow,1/2}|^2. \end{equation}
Using this equation and the fact that wavefunction is normalized (assuming the $J_z=\pm1$ components are zero), we can write Eq.~(\ref{rho_S}) as 
\begin{equation} \label{rho_Si_sigma} 
\rho_{I}=\frac{1}{2}\Big[\textbf{1} + \frac{d}{v_0} j_p \sigma_z \Big]. 
\end{equation}
The probability current vanishes in the energy gaps, which therefore implies that the entanglement entropy, $S=-\hbox{Tr}\rho_I\log\rho_I$, reaches its maximum value of $\log2$ at the band edges~\footnote{At the band edges (from Eq.~(\ref{cond1/2}), the current vanishes continuously as well.}. This finding is consistent with numerical results, as demonstrated in Fig.~\ref{fig-entropy}. One would expect that the spectral gaps occur at values of the magnetic flux where the electronic and impurity spins interact most strongly, and the fact that the entanglement is greatest near these values is consistent with this picture.


\section{Origin of flat bands and parameter dependence for spin 1/2 impurity} \label{sec_gaps}

\begin{figure}
\includegraphics[width=\columnwidth]{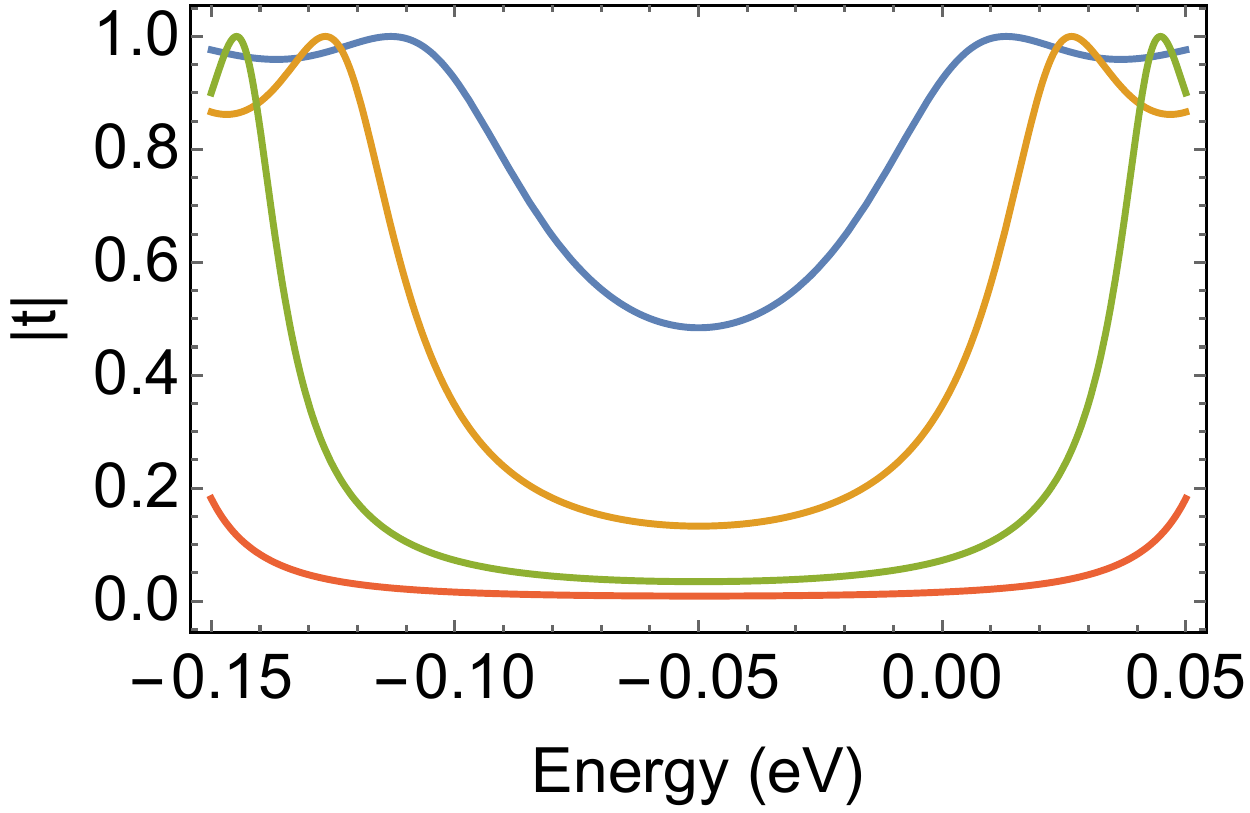}
\caption{(Color online)~Magnitude of the transmission coefficient as a function of energy for a single spin 1/2 impurity with square potential for several different values of $A^\perp$. From top to bottom: $A^\perp=0.025, 0.05,0.075,0.1$ eV. The remaining parameters are $d=1000$ \AA, $w=130$ \AA, $A^z=0.05~\text{eV}$ \cite{Liu}, $v_0=2.4~\text{eV \AA}$.}
\label{fig-abst}
\end{figure}

In this section, we focus on the case of a single spin 1/2 impurity with a square potential, and we investigate quantitatively how the spectrum depends on the system parameters. As we discussed in Sections~\ref{sec_half} and \ref{sec_integer}, the energy spectrum is completely determined by the transmission amplitude $t$ (see Eq.~\eqref{lambda1/2}). As we show in Appendix~\ref{appendix_general_rt}, for a square potential this coefficient can be written as,
\begin{equation} \label{t_coefficient}
t(E)= \frac{e^{iE(d-w)/v_0}v_0q}{v_0q\cos(qw)-i(A^z+E)\sin(qw)},  
\end{equation}
where $v_0q=\sqrt{(E+A^z)^2-(A^\perp)^2}$. The magnitude of the transmission amplitude is minimal at $E=-A^z$, at which it assumes the value $|t|=\hbox{sech}(wA^\perp/v_0)$. Thus, the transmission is exponentially suppressed as the width $w$ or height $A^\perp$ of the impurity barrier are increased, or as the electron velocity $v_0$ is reduced. The dependence of $|t|$ on $A^\perp$ is demonstrated in Fig.~\ref{fig-abst}, where it is evident that $|t|$ flattens out close to zero over a broad range of energies that grows as $A^\perp$ is increased. This behavior gives rise to the flat bands that occur in the middle of Fig.~\ref{fig-spin-half-band}(a). To see this, recall that the condition for a band to occur is $|\cos\phi_t|<|t|$, so that for $|t|\ll1$, the range of phases satisfying this condition becomes very narrow. For physical parameters, $\phi_t$ is dominated by the kinematic term $E(d-w)/v_0$, which is why the parametric plot trajectories shown in Figs.~\ref{fig-spin-half-band}(b) and \ref{fig-spin-one-band}(b) are nearly circular. Here, we can use this observation to estimate the smallest bandwidth $\sigma_E$ and the largest bandgap $\Delta E$, which occur near $E=-A^z$: $\sigma_E\approx2v_0/(d-w)\arcsin(\hbox{sech}(wA^\perp/v_0))$, $\Delta E\approx\pi v_0/(d-w)$. Notice that $\sigma_E$ depends sensitively on the impurity coupling while $\Delta E$ does not. For the typical experimental parameters used in Fig.~\ref{fig-spin-half-band}, these quantities evaluate to $\sigma_E\approx0.7~\text{meV}$ and $\Delta E\approx8~\text{meV}$, corresponding to the flat bands in the vicinity of $E=-0.05$ eV in Fig.~\ref{fig-spin-half-band}(a).

\begin{figure*}%
    \centering
   {{\includegraphics[scale=.721]{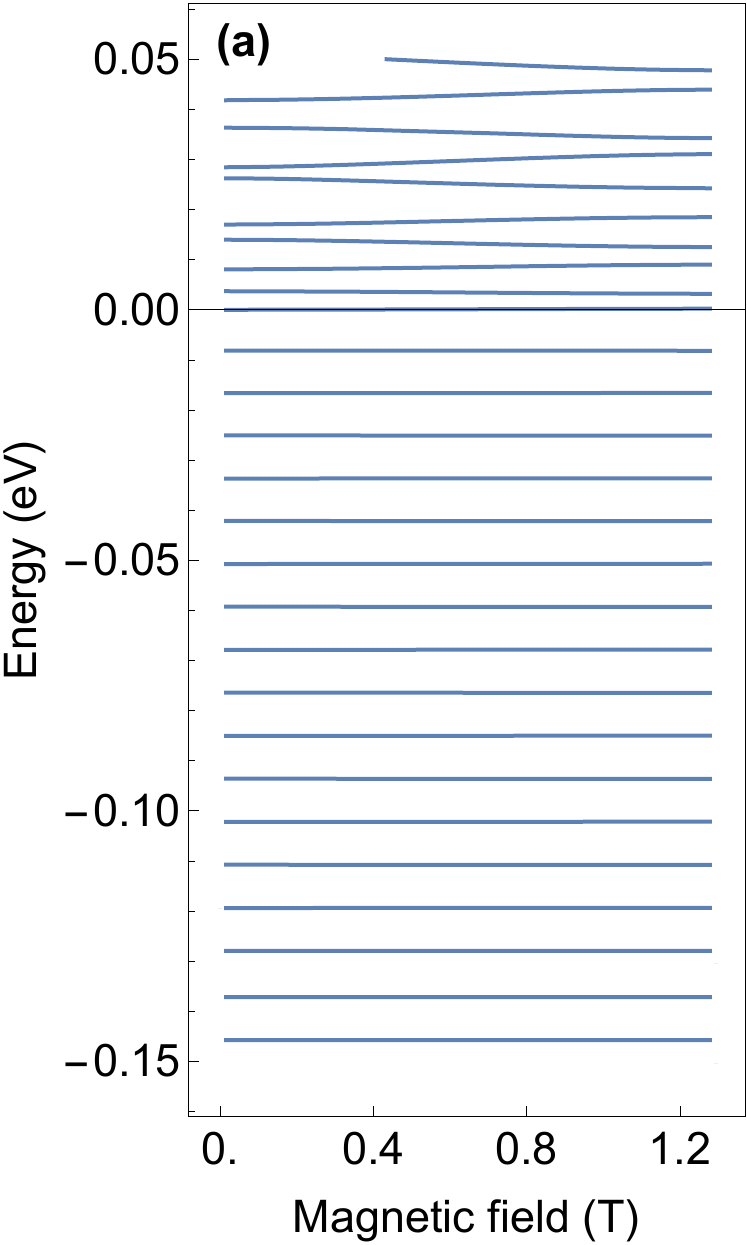} }}
   \hspace{-0.34cm}
   {{\includegraphics[scale=.721]{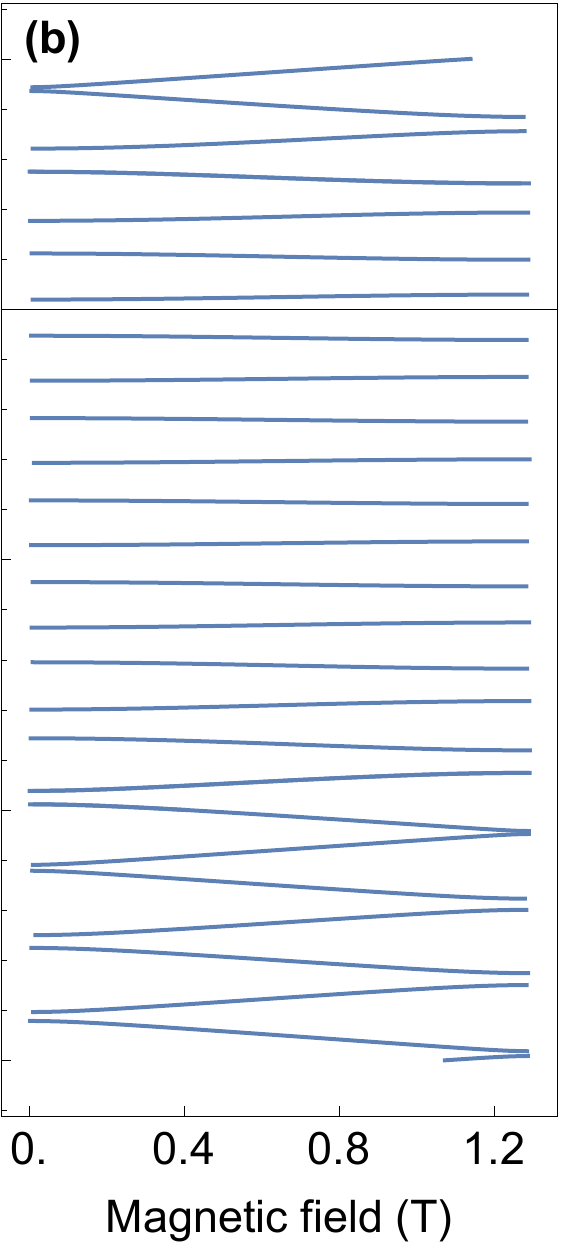} }}\hspace{-0.27cm}
   {{\includegraphics[scale=.72]{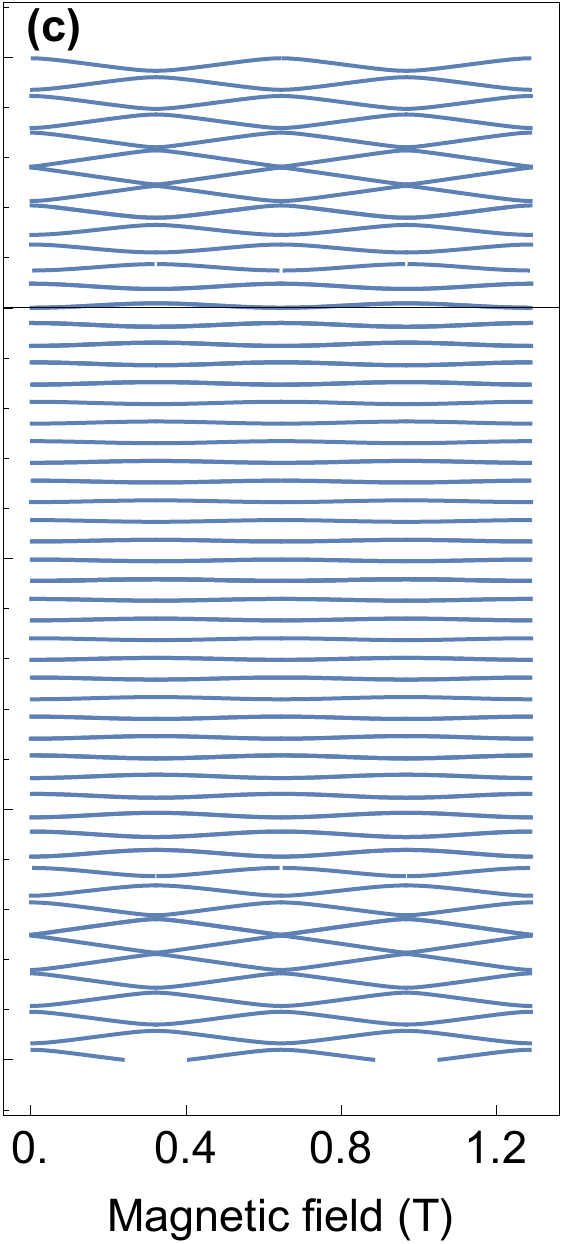} }}\hspace{-0.27cm}
   {{\includegraphics[scale=.72]{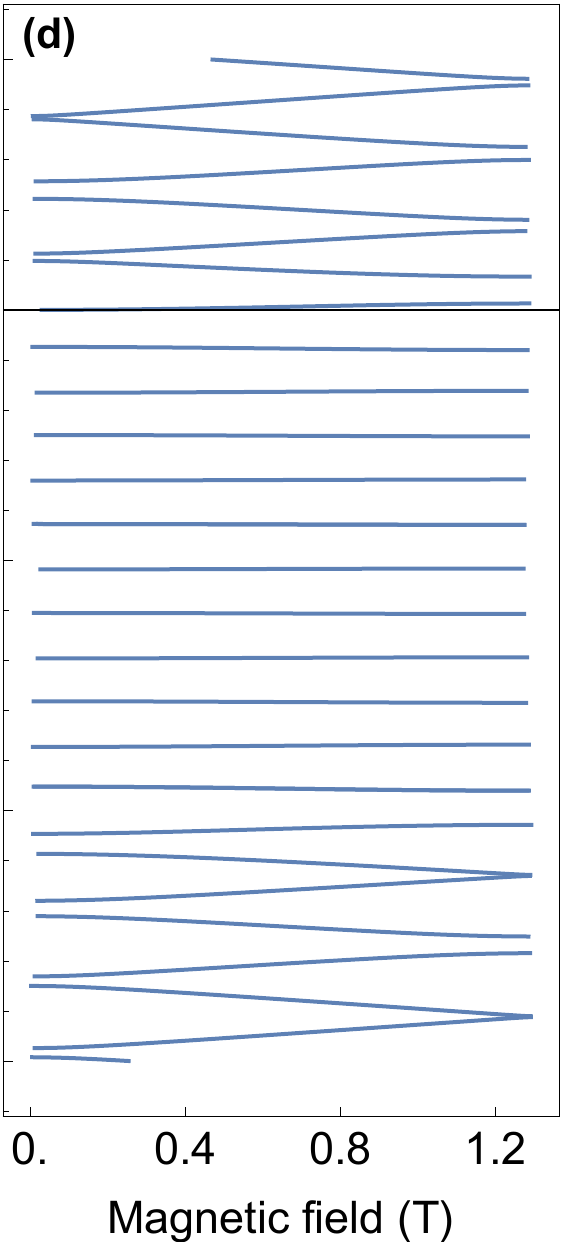} }}
     \caption{(Color online)~Dependence of energy spectrum on ring geometry and impurity couplings for a spin 1/2 impurity with square potential. The parameters are (a) $d=1000$ \AA, $w=130$ \AA, $A^z=A^\perp=0.05~\text{eV}$ (b) $d=1000$ \AA, $w=130$ \AA, $A^z=0.03,~A^\perp=0.05~\text{eV}$ (c) $d=2000$ \AA, $w=130$ \AA, $A^z=A^\perp=0.05~\text{eV}$  and (d) $d=1000$ \AA, $w=200$ \AA, $A^z=A^\perp=0.05~\text{eV}$. In all cases, $v_0=2.4~\text{eV \AA}$. For the larger ring circumference case shown in (c), the magnetic field range spans several Brillouin zones.}%
     \label{fig-compare}%
\end{figure*}

To further elucidate the dependence on system parameters, we show the energy spectrum for several sets of parameters in Fig.~\ref{fig-compare}. In Fig.~\ref{fig-compare}(a), we increase the strength of the impurity coupling by an order of magnitude relative to Fig.~\ref{fig-spin-half-band}, where the most striking consequence is that the bands for $E<0$ become significantly flatter ($\sigma_E\sim10^{-14}$ eV near $E=-A^z=-0.5$ eV in this case). Notice that the band gaps remain approximately the same, as is consistent with our finding that these are insensitive to the impurity coupling. In addition, the flat bands continue over a 1 eV range in this case, all the way down to $E\approx-1$ eV, since now $A^\perp=0.5$ eV. 

In Fig.~\ref{fig-compare}(b), we consider a situation in which $A^z\ne A^\perp$. In particular, we keep $A^\perp=0.05$ eV as in Fig.~\ref{fig-spin-half-band}, but now reduce the longitudinal coupling to $A^z=0.03$ eV. This shifts the flat band region upward in energy but does not affect $\sigma_E$ or $\Delta E$. An additional consequence of $A^z\ne A^\perp$ is that the behavior near $E=0$ is modified. When $A^z=A^\perp$ and $E=0$, it follows from Eq.~\eqref{t_coefficient} that $t=v_0/(v_0-i A^\perp w)$, which saturates the band condition $|\cos(\phi_t)|=|t|$. Thus, $E=0$ always corresponds to a band edge in the case of an isotropic interaction, while this property is lost in the anisotropic case. This behavior is evident from a comparison of Fig.~\ref{fig-compare}(b) with the other panels of that figure.

Figs.~\ref{fig-compare}(c),(d) show the dependence of the spectrum on the geometry of the ring. Increasing the ring circumference changes the number of ``Brillouin zones'' that fit within a given range of magnetic field, as shown in Fig.~\ref{fig-compare}(c). Here, we increase the circumference by a factor of 2 relative to Fig.~\ref{fig-spin-half-band}, so that now two full Brillouin zones fit instead of only half of one. In addition, the density of states increases by a factor of 2, as follows from the inverse dependence of $\sigma_E$ and $\Delta E$ on the circumference.

The spectra shown in Fig.~\ref{fig-compare} contain a number of very small gaps, raising the question of whether the gaps ever close completely to form a Dirac point. As was mentioned in Sec.~\ref{sec_half}, band touching points can arise if the band edge condition, $|\cos\phi_t|=|t|$, and the transparency condition, $|t|=1$, are simultaneously satisfied. In the case of a square impurity barrier, we see from Eq.~\eqref{t_coefficient} that $|t|=1$ when $q=n\pi/w$ for arbitrary nonzero integer $n$, which corresponds to the following energies
\begin{equation}
E_n^\pm=-A^z\pm\sqrt{(A^\perp)^2+n^2\pi^2v_0^2/w^2}.
\end{equation}
At these energies, the transmission amplitude reduces to a pure phase:
\begin{equation}
t(E_n^\pm)=\pm e^{iE_n^\pm(d-w)/v_0}.
\end{equation}
Notice that these energies are always guaranteed to lie within a band since $|\cos\phi_t|\le|t|$ is automatically satisfied. Imposing the band edge condition, $\phi_t=m\pi$ for integer $m$, then leads to the following set of discrete values of the ring circumference for which Dirac points appear in the spectrum:
\begin{equation}\label{dvals}
d=w+m\pi v_0/E_n^\pm.
\end{equation}
Any choice of $m$ will yield a Dirac point at energy $E_n^\pm$. Although the particular form of Eq.~\eqref{dvals} only holds in the idealized case of a square impurity potential, an analogous expression should arise for other potential shapes.


\section{Conclusion}

In conclusion, we analyzed the problem of a topological insulator ring in which the helical edge states are coupled to magnetic impurities or spinful nuclei of arbitrary spin. This interaction breaks time-reversal symmetry and enables the backscattering of electrons. We considered the case where the ring is threaded by a magnetic flux, and we showed that the energy spectrum as a function of this flux is given by a universal formula that depends only on the amplitude of transmission through the impurity. We found that the impurity can give rise to sizable spectral gaps and flat bands, and we calculated the gap sizes and bandwidths for a variety of experimentally relevant parameter regimes. We further showed that the entanglement between the electronic and impurity spins is maximal near these gaps, while at energies far away from these gaps, little entanglement develops, and the helical edge states remain unaffected by the impurity. Our results can be tested with quantum interference measurements in nanorings, providing a new approach to understanding the role of magnetic impurities in topological insulator transport.




\appendix



\section{Generalized time-reversal relations for spin 1/2 impurities} \label{1/2_t_r_appendix}

GTR symmetry mixes eigenstates incoming from the left of the impurity with those incoming from the right. In this appendix, we exploit this fact to obtain a simple expression for the eigenstate spectrum in terms of the scattering transmission amplitude. This result is universal in the sense that it does not depend on the spatial profile (barrier shape) of the impurity or any other details of the system. We begin by supposing that the initial impurity state is an eigenstate of $I_z$, and we write the wavefunctions in terms of the transmission and reflection amplitudes. For example, we denote the left-incoming scattering eigenstate with initial impurity state $\ket{-1/2}$ by $\ket{\phi_{\rightarrow,-1/2}^{(p)}}$ (total $J_z=0$ and momentum $p$). This state corresponds to the $A$ wave in Fig.~\ref{fig-Incidents}. The state on each side of the impurity takes the form ($x$ denoting left side of the barrier and $x'$ the right side)
\begin{equation} \label{A_0}
\langle x \ket{\phi_{\rightarrow,-1/2}^{(p)}}=\begin{bmatrix} 0 \\ e^{i p x} \\ r_{\rightarrow}e^{-i p x} \\ 0\end{bmatrix}, \quad \langle x' \ket{\phi_{\rightarrow,-1/2}^{(p)}}=\begin{bmatrix}0 \\ t_{\rightarrow}e^{i p x'} \\ 0 \\ 0\end{bmatrix}.
\end{equation}
Similarly, for the $B$ wave (incident from the right with the impurity initially in $\ket{1/2}$) we have,
\begin{equation} \label{B_0}
\langle x \ket{\phi_{\leftarrow,1/2}^{(p)}}=\begin{bmatrix} 0 \\ 0 \\ t_{\leftarrow}e^{-i p x} \\ 0\end{bmatrix}, \quad \langle x' \ket{\phi_{\leftarrow,1/2}^{(p)}}=\begin{bmatrix}0 \\ r_{\leftarrow}e^{i p x'} \\ e^{-i p x'} \\ 0\end{bmatrix}.\end{equation}
We also have an eigenstate corresponding to a left-incoming electron with the impurity initially in state $\ket{1/2}$:
\begin{equation} 
\label{Ap_0}  \langle x \ket{\phi_{\rightarrow,1/2}^{(p)}}=\begin{bmatrix}  e^{i p x} \\ 0 \\0 \\ 0\end{bmatrix}, \quad  \langle x'\ket{\phi_{\rightarrow,1/2}^{(p)}}=\begin{bmatrix}\mathcal{P}_\rightarrow e^{i p x'} \\ 0 \\ 0\\ 0\end{bmatrix},
\end{equation}
and similarly for $\ket{\phi_{\leftarrow,-1/2}^{(p)}}.$ 

We define the TR operator as $\mathcal{T}_{1/2}=i\sigma_y\Theta$ where $\Theta$ is the complex conjugation operator. The corresponding operator for the GTR symmetry in the case of a spin 1/2 impurity is then $\mathcal{T}_{GTR}=\mathcal{T}_{1/2} \otimes \mathcal{T}_{1/2}$. It is easy to see that $\mathcal{T}_{GTR}\ket{\phi_{\rightarrow,1/2}^{(p)}}$ must be proportional to $\ket{\phi_{\leftarrow,-1/2}^{(p)}},$ from which we conclude that for ``passing'' states like Eq.~\eqref{Ap_0},
\begin{equation}
\mathcal{P}_\rightarrow=\mathcal{P}_\leftarrow\equiv\mathcal{P} \quad \text{with} \quad |\mathcal{P}|^2=1.
\end{equation}
Applying $\mathcal{T}_{GTR}$ on the two states with $J_z=0$, Eqs.~\eqref{A_0} and \eqref{B_0}, we see that the resulting state is a superposition of left-incoming and right-incoming states: 
\begin{equation} 
\label{T_A_0 }\mathcal{T}_{GTR} \ket{\phi_{\rightarrow,-1/2}^{(p)}}=a_0\ket{\phi_{\rightarrow,-1/2}^{(p)}}+b_0\ket{\phi_{\leftarrow,1/2}^{(p)}}.
\end{equation}
For instance, $\mathcal{T}_{GTR} \ket{\phi_{\rightarrow,-1/2}^{(p)}}$ corresponds to the following left-side and right-side wavefunctions:
\begin{eqnarray}
 \label{T_example } \langle x|\mathcal{T}_{GTR} \ket{\phi_{\rightarrow,-1/2}^{(p)}}&=&\begin{bmatrix} 0 \\  -\bar{r}_{\rightarrow}e^{i p x}\\ -e^{-i p x} \\ 0\end{bmatrix},\nonumber\\ \langle x'|\mathcal{T}_{GTR}\ket{\phi_{\rightarrow,-1/2}^{(p)}}&=&\begin{bmatrix}0\\ 0\\ -\bar{t}_{\rightarrow}e^{-i p x'} \\ 0\end{bmatrix}.
\end{eqnarray}
Imposing Eq.~(\ref{T_A_0 }) to the wavefunction on the left side of the barrier will give us one equation per component. One of these equations implies that $a_0=-\bar{r}_{\rightarrow}$. We do the same on the right side, and the equation resulting from the third component implies $b_0=-\bar{t}_{\rightarrow}.$ The remaining components (applied on both left and right) provide the following equations:
\begin{eqnarray}
&&\bar{r}_{\rightarrow}t_{\rightarrow}+r_{\leftarrow}\bar{t}_{\rightarrow}=0,\label{ampeqn1}\label{ampeqn1} \\ &&|r_\rightarrow|^2 +t_{\leftarrow}\bar{t}_{\rightarrow}=1.\label{ampeqn2}
\end{eqnarray}
Combining Eq.~\eqref{ampeqn2} with the current conservation condition, $|r_{\rightarrow}|^2+|t_{\rightarrow}|^2=1$, we find that the two transmission amplitudes must be equal:
\begin{equation}
t_{\rightarrow}=t_{\leftarrow}\equiv t\equiv|t|e^{i \phi_t}.\label{ampeqn3}
\end{equation}
Multiplying Eq.~\eqref{ampeqn1} by $r_{\rightarrow}t$ and simplifying the result with the help of \eqref{ampeqn2} and \eqref{ampeqn3}, we also find
\begin{equation}
r_{\leftarrow}r_{\rightarrow}=(|t|^2-1)e^{2i\phi_t}.\label{ampeqn4}
\end{equation}
Eqs.~\eqref{ampeqn3} and \eqref{ampeqn4} are used to derive the remarkably simple expression for the band structure given in Eq.~\eqref{lambda1/2}.

So far the only assumption we have made is that the Hamiltonian has GTR symmetry. If the impurity potential also possesses inversion symmetry (as a result of which $r_\rightarrow=r_\leftarrow\equiv r\equiv|r|e^{i\phi_r}$~\footnote{Note that $t_\rightarrow=t_\leftarrow$ holds regardless.}), then Eq.~\eqref{ampeqn1} implies $\bar{r}t = - r\bar{t},$ which in turn leads to $\phi_t=\phi_r+\pi/2$. The presence of this symmetry of course has no impact on Eq.~\eqref{lambda1/2}, but it does simplify the calculation of the wavefunction, for example in the case of a square barrier considered in Appendix~\ref{appendix_general_rt}.


\section{Transmission and reflection amplitudes for square impurity barrier} \label{appendix_general_rt}

In this appendix, we outline the general approach for finding reflection and transmission amplitudes for an arbitrary-spin impurity with square potential (i.e., $F(y)=\Theta(w/2-|y|)$ in Eq.~\eqref{hyperfine}). We do this by decomposing the Hilbert space into sectors of total spin $J_z$ and by separately solving for the scattering amplitudes in each sector. To better understand the structure of the Hamiltonian in each sector, we first consider the case of a spin 1/2 impurity. Using a wave function ansatz that includes a chiral plane wave factor of the form $e^{iqy}$, the full Hamiltonian $H=H_0+H_{S,I}$ inside the interaction region is
\begin{equation} \label{s1/2d} 
H=\begin{bmatrix} A^z+q v_0 & 0 & 0 & 0 \\0 & q v_0-A^z & A^\perp & 0 \\0 & A^\perp & -q v_0-A^z & 0 \\0 & 0 & 0 & A^z-q v_0 \end{bmatrix}. 
\end{equation}
The middle block of this matrix corresponds to the $J_z=0$ sector ($M_{(J_z=0)}$) and the other two states are $\ket{\uparrow\uparrow}$ and $\ket{\downarrow\downarrow}$, which do not couple to any other states and are irrelevant for calculating scattering amplitudes. For a general impurity spin, $H$ will be block diagonal (when the basis states are grouped according to $J_z$), all of which are two-dimensional except for the one-dimensional blocks corresponding to maximal $|J_z|$. Each two-dimensional block, which we denote by $M_{(J_z)}$, will have an associated set of scattering amplitudes $r_\leftarrow$, $r_\rightarrow$, $t_\leftarrow$, $t_\rightarrow$. The most general form of $M_{(J_z)}$ for arbitrary $J_z$ is
 \begin{equation} \label{Bq} 
 \begin{split}
M_{(J_z)}=\begin{bmatrix} qv_0 - u -m_0 & h \\ h & -qv_0-u+m_0 \end{bmatrix} \\ = -u \textbf{1} + (qv_0-m_0) \sigma_z + h \sigma_x \\ 
=-u \textbf{1} + [b\cos\theta \sigma_z + b\sin\theta \sigma_x],\end{split} 
\end{equation}
where we have defined
\begin{equation} \label{Bq_relation} 
\begin{bmatrix} \cos\theta  \\  \sin\theta \end{bmatrix} =  \frac{1}{b}\begin{bmatrix} qv_0-m_0  \\  h \end{bmatrix} \quad \text{and} \quad b^2=h^2+(qv_0-m_0)^2. 
\end{equation}
The eigenvectors are 
\begin{equation} \label{Bq_vectors} 
\begin{bmatrix} \cos\theta/2  \\  \sin\theta/2 \end{bmatrix} \quad \text{and} \quad  \begin{bmatrix} -\sin\theta/2 \\ \cos\theta/2   \end{bmatrix}, 
\end{equation}
and the eigenvalues are $E=-u \pm b$. We can solve for $q$ in terms of energy,
\begin{equation} \label{Bq_q} 
q_\pm=(1/v_0)\bigg( m_0 \pm \sqrt{(E+u)^2 - h^2}   \bigg).
\end{equation}
Note that $q$ and hence $\theta$ may be complex depending on the energy.

To obtain the scattering amplitudes, we need to match wavefunction ansatzes inside and outside the impurity potential at the boundaries of the potential. Defining $a=d-w$, we match the wavefunctions at $y=a/2$ and $y=a/2+w$ (see Fig.~\ref{fig-Incidents}), where 
\begin{equation} \label{Bq_wf} 
 \psi(0)=\begin{bmatrix} A \\ r_\rightarrow A+t_\leftarrow B \end{bmatrix},\quad  \psi(d)=\begin{bmatrix} t_\rightarrow A+r_\leftarrow B \\ B \end{bmatrix}. 
\end{equation}
On the left side of the interaction region, this amounts to requiring
\begin{equation}
\begin{bmatrix} Ae^{ipa/2} \\ (r_\rightarrow A+t_\leftarrow B)e^{-ipa/2} \end{bmatrix} = \begin{bmatrix} \cos\theta/2 \\ \sin\theta/2 \end{bmatrix} c_+  + \begin{bmatrix}  \sin\theta/2 \\ \cos\theta/2 \end{bmatrix} c_- ,
\end{equation}
and on the right side,
\begin{eqnarray}
&&\begin{bmatrix} (t_\rightarrow A+r_\leftarrow B)e^{-ipa/2} \\ Be^{ipa/2} \end{bmatrix} =\nonumber\\&&  \begin{bmatrix} \cos\theta/2 \\ \sin\theta/2 \end{bmatrix} e^{i q_+ w} c_+  + \begin{bmatrix}  \sin\theta/2 \\ \cos\theta/2 \end{bmatrix} e^{i q_- w}  c_-,
\end{eqnarray}
where $p=E/v_0$. These equations, combined with the single-valuedness condition \eqref{singlevalued}, allow us to eliminate $A$, $B$, and $c_\pm$ and to obtain the reflection and transmission amplitudes.
As an example, for the case of spin 1/2 impurity where $m_0=0$ and $q_\pm = \pm q$, we find 
\begin{eqnarray}
\!\!\!\!\!\!\sqrt{r_\leftarrow r_\rightarrow} &=& i e^{i p a} \frac{\sin\theta \sin q w}{e^{- i q w } \cos^2 \theta/2 - e^{i q w } \sin^2\theta/2},\\  
\!\!\!\!\!\!t_\leftarrow=t_\rightarrow&=& e^{i p a} \frac{\cos\theta}{e^{- i q w } \cos^2 \theta/2 - e^{i q w } \sin^2\theta/2}. 
\end{eqnarray}
With some simple algebraic manipulations, we can transform this transmission amplitude to the form of Eq.~(\ref{t_coefficient}). In addition, we may write 
\begin{eqnarray}
|r|^2&=&\frac{\sin^2 (q w) \sin^2\theta}{1-\cos^2(q w) \sin^2 \theta}
= \frac{\sin^2 (q w)}{1+\cot^2\theta - \cos^2(qw)} \nonumber\\ &=& \bigg[ 1+\frac{\cot^2\theta}{\sin^2((h/v_0)w\cot\theta)}    \bigg]^{-1},
\end{eqnarray}
where we have used 
\begin{equation} \label{csc} 
\sin\theta = \frac{h}{E+u}=\frac{A^\perp}{E+A^z},  
\end{equation}
where the first equality holds for any sector, while the second equality applies for $J_z=0$. Note that we can express all other variables in terms of this shifted (and dimensionless) energy $(E+u)/h$. For example, we may rewrite the transmission amplitude as
\begin{equation}
t= \frac{e^{iE(d-w)/v_0}v_0q}{v_0q\cos(qw)-i(A^z+E)\sin(qw)}.
\end{equation}


\section{Generalized time-reversal relations for spin 1 impurities} \label{1_t_r_appendix}

In this appendix, we derive the consequences of GTR symmetry in the case of a spin 1 impurity. As explained in Section~\ref{sec_overview}, the wavefunction decomposes into blocks spanned by basis states with the same absolute value of total spin $|J_z|$. As discussed in Sec.~\ref{sec_integer}, the spectrum for an arbitrary-spin impurity can be obtained by combining the solutions for spin 1/2 (obtained in Sec.~\ref{sec_half}) and spin 1 impurities. In the case of a spin 1 impurity, the electron-impurity Hilbert space divides into two trivial one-dimensional subspaces corresponding to the states with $J_z=\pm3/2$, and a four-dimensional block spanned by states with $J_z=\pm 1/2$. Following the procedure of Appendix~\ref{1/2_t_r_appendix}, we consider left-incoming and right-incoming states for which the impurity is initially in an eigenstate of $I_z$. We label these states as e.g., $\ket{\phi^{(p)}_{\rightarrow,0}}$, which represents an electron incoming from the left with momentum $p$ and with the impurity initially in state $\ket{0}$. We write the wavefunctions for each of these states on the left-side ($x$) and right-side ($x'$) of the impurity in terms of reflection and transmission amplitudes:
\begin{equation}
\langle x\ket{\phi_{\rightarrow,0}^{(p)}}=\begin{bmatrix} e^{i p x} \\ 0 \\ r_{\rightarrow}^+e^{-i p x} \\ 0\end{bmatrix},\quad  \langle x'\ket{\phi_{\rightarrow,0}^{(p)}}=\begin{bmatrix} t_{\rightarrow}^+e^{i p x'} \\ 0 \\ 0\\ 0\end{bmatrix},\label{phiright0}
\end{equation}
\begin{equation}
\langle x\ket{\phi_{\rightarrow,-1}^{(p)}}=\begin{bmatrix}0\\ e^{i p x} \\ 0 \\ r_{\rightarrow}^-e^{-i p x} \end{bmatrix},\quad  \langle x'\ket{\phi_{\rightarrow,-1}^{(p)}}=\begin{bmatrix} 0\\ t_{\rightarrow}^-e^{i p x'} \\ 0 \\ 0\end{bmatrix},\label{phirightm}
\end{equation}
\begin{equation}
\langle x\ket{\phi_{\leftarrow,1}^{(p)}}=\begin{bmatrix} 0 \\ 0 \\ t_{\leftarrow}^+e^{-i p x} \\ 0\end{bmatrix},\quad  \langle x'\ket{\phi_{\leftarrow,1}^{(p)}}=\begin{bmatrix} r_{\leftarrow}^+e^{i p x'} \\ 0 \\ e^{-i p x'}\\ 0\end{bmatrix},\label{phileftp}
\end{equation}
\begin{equation}
\langle x\ket{\phi_{\leftarrow,0}^{(p)}}=\begin{bmatrix}0\\ 0 \\ 0 \\ t_{\leftarrow}^-e^{-i p x} \end{bmatrix},\quad  \langle x'\ket{\phi_{\leftarrow,0}^{(p)}}=\begin{bmatrix}0\\ r_{\leftarrow}^-e^{i p x'} \\ 0 \\ e^{-i p x'}\end{bmatrix}.\label{phileft0}
\end{equation}
Here, the basis states are $\ket{\uparrow,0}$, $\ket{\uparrow,-1}$, $\ket{\downarrow,0}$, $\ket{\downarrow,1}$; we have left out the two ``passing'' states $\ket{\uparrow,1}$ and $\ket{\downarrow,-1}$ since the action of GTR on these will be identical to that for the passing states in the spin 1/2 case treated in Appendix~\ref{1/2_t_r_appendix}, namely the impurity-induced phases on these states obey the relation $\mathcal{P}_\rightarrow=\mathcal{P}_\leftarrow=\mathcal{P}$. 

In order to understand the action of GTR on these states, we must first generalize the definition of the GTR operator introduced in Appendix~\ref{1/2_t_r_appendix} to the case of a spin 1 impurity. We choose the following definition:
\begin{equation} 
\label{trvr_1} \mathcal{T}_{1}=\begin{pmatrix}  0 & 0 & 1 \\ 0 & -1 & 0 \\ 1 & 0 & 0 \end{pmatrix}\Theta, 
\end{equation}
where $\Theta$ is again the complex conjugation operator. The GTR operator is then $\mathcal{T}_{GTR}=\mathcal{T}_{1/2} \otimes \mathcal{T}_{1}$. Acting with this operator on one of the states in Eqs.~\eqref{phiright0} - \eqref{phileft0} yields a linear combination of two of the other states. For example,
\begin{equation} \label{T_A_p}
\mathcal{T}_{GTR} \ket{\phi_{\rightarrow,0}^{(p)}} =a^-\ket{\phi_{\rightarrow,-1}^{(p)}} + b^-\ket{\phi_{\leftarrow,0}^{(p)}},
\end{equation}
and
\begin{equation} \label{T_A_m}
\mathcal{T}_{GTR}\ket{\phi_{\rightarrow,-1}^{(p)}} =a^+\ket{\phi_{\rightarrow,0}^{(p)}} + b^+\ket{\phi_{\leftarrow,1}^{(p)}}.
\end{equation}
By acting on the other two states in a similar fashion, we get a total of four equations like Eqs.~\eqref{T_A_p} and \eqref{T_A_m}, each of which yields two 4-component spinor equations when we restrict to the left- or right-side of the impurity. This gives a total of 32 complex equations, 16 of which are trivial, and 8 more can be used to solve for the 8 coefficients $a^-$, $b^-$, etc. The remaining 8 complex equations constrain the scattering amplitudes and can be written as
\begin{enumerate}[label=(\roman*)]
\centering
\item \label{1} $1= \bar{t}_{\rightarrow}^+ t_{\leftarrow}^-+\bar{r}_{\rightarrow}^+ r_{\rightarrow}^-$
\item \label{2}$\bar{t}_{\rightarrow}^+ r_{\leftarrow}^- = -\bar{r}_{\rightarrow}^+t_{\rightarrow}^-$ \\
\item \label{3}$1 = \bar{t}_{\rightarrow}^-  t_{\leftarrow}^++\bar{r}_{\rightarrow}^- r_{\rightarrow}^+$
\item \label{4}$\bar{r}_{\rightarrow}^- t_{\rightarrow}^+ =-\bar{t}_{\rightarrow}^-  r_{\leftarrow}^+$ \\

\item \label{5}$\bar{t}_{\leftarrow}^+r_{\rightarrow}^- = -\bar{r}_{\leftarrow}^+t_{\leftarrow}^-$
\item \label{6}$1=  \bar{t}_{\leftarrow}^+t_{\rightarrow}^- +\bar{r}_{\leftarrow}^+r_{\leftarrow}^-$\\

\item \label{7}$\bar{t}_{\leftarrow}^- r_{\rightarrow}^+ = -\bar{r}_{\leftarrow}^- t_{\leftarrow}^+  $
\item \label{8}$1 = \bar{t}_{\leftarrow}^-t_{\rightarrow}^++\bar{r}_{\leftarrow}^-r_{\leftarrow}^+ $
\end{enumerate}
Equations (i) and (ii) come from solving Eq.~\eqref{T_A_p} on the left- and right-side of the impurity, equations (iii) and (iv) come from solving Eq.~\eqref{T_A_m}, and equations (v), (vi) and (vii), (viii) come from solving similar equations involving $\mathcal{T}_{GTR}\ket{\phi_{\leftarrow1}}$ and $\mathcal{T}_{GTR}\ket{\phi_{\leftarrow0}}$, respectively. An instant consequence of these eight equations is,

\begin{eqnarray}
\label{ts}   &&|t_{\rightarrow}^-|=|t_{\rightarrow}^+|= |t_{\leftarrow}^+|= |t_{\leftarrow}^-|= |t|,\\
\label{rs} && |r_{\rightarrow}^-|=|r_{\rightarrow}^+|= |r_{\leftarrow}^+|= |r_{\leftarrow}^-|= |r|.
\end{eqnarray}
Applying this constraint to equations \ref{2}, \ref{4}, \ref{6} and \ref{8} gives a relation between the phases. Adding the resulting equations from \ref{2} and \ref{8} gives
\begin{equation} \label{t_s_1}  
\phi_{t_\leftarrow^+}- \phi_{t_\rightarrow^+}= \phi_{t_\rightarrow^-}-\phi_{t_\leftarrow^-}, \end{equation}
and adding the resulting equations from \ref{6} and \ref{8} gives
\begin{equation}  
\phi_{r_\rightarrow^+}-\phi_{r_\leftarrow^+}= \phi_{r_\rightarrow^-}- \phi_{r_\leftarrow^-}. \end{equation}
Furthermore, solving the all eight equations with the constraint that $|t|^2+|r|^2=1$, provides us with
\begin{eqnarray}   
&&\phi_{t_\rightarrow^-} = \phi_{t_\leftarrow^+}=\phi_{r_\rightarrow^-}+\phi_{r_\leftarrow^+}-\phi_{t_\leftarrow^-}+\pi\nonumber \\ && \phi_{t_\leftarrow^-}=\phi_{t_\rightarrow^+}\nonumber\\  &&\phi_{r_\rightarrow^+}=\phi_{r_\rightarrow^-}\nonumber \\ &&\phi_{r_\leftarrow^-}=\phi_{r_\leftarrow^+}\label{8_solved}
\end{eqnarray}
from which Eqs.~(\ref{rtlr1}) and (\ref{rtwf1}) follow.

Now we proceed to derive Eq.~(\ref{lambda1}) from the equation that results from imposing the single-valuedness condition, Eq.~\eqref{singlevalued}, in each sector:
\begin{eqnarray} \label{gamma1}
&&e^{2\pi i \Phi_B/\Phi_0}=\nonumber\\&& \!\!\!\frac{1- r^\pm_{\leftarrow}r^\pm_{\rightarrow}+t^\pm_{\leftarrow} t^\pm_{\rightarrow} \pm \sqrt{(1- r^\pm_{\leftarrow}r^\pm_{\rightarrow}+t^\pm_{\leftarrow} t^\pm_{\rightarrow})^2 -4t^\pm_{\leftarrow} t^\pm_{\rightarrow}} }{2  t^\pm_{\leftarrow}}.\nonumber\\&&
\end{eqnarray}
Here, the $\pm$ in front of the square root is independent from the sector labels $\pm$ labeling the scattering amplitudes. Introducing the average amplitude phases as in Sec.~\ref{sec_integer}, $\phi_x^\circ=(\phi_{x\rightarrow}+\phi_{x\leftarrow})/2$ where $x=r^\pm,t^\pm$, we can rewrite the above expression using Eqs.~\eqref{ts}, \eqref{rs}, \eqref{8_solved}:
\begin{equation} \label{global}
e^{2\pi i \tfrac{\Phi_B}{\Phi_0}}=e^{i (\phi_{t^\pm_\rightarrow}- \phi_{t^\pm_\leftarrow})/2}\left[\cos\phi_t^\circ/|t| \pm \sqrt{(\cos\phi_t^\circ/|t|)^2-1}\right], 
\end{equation}
where $\phi_{t^+}^\circ=\phi_{t^-}^\circ\equiv\phi_t^\circ$. From Eq.~\eqref{t_s_1} we see that the overall phase factor differs only by a sign between the two sectors, $\phi_{t^+_\rightarrow}- \phi_{t^+_\leftarrow}=-(\phi_{t^-_\rightarrow}- \phi_{t^-_\leftarrow})$. Combined with the two possible branches of the square root in Eq.~\eqref{global}, this therefore produces four distinct values of $\Phi_B$ for each value of the energy.

Next, we show how to derive the ratio of spin and probability currents given in Eq.~(\ref{jsp1}). First, we apply the single-valuedness condition, Eq.~\eqref{singlevalued}, to Eq.~(\ref{rtwf1}) to obtain a formula for the wavefunction coefficients:
\begin{equation} \label{Gammas}  
\frac{B^\pm}{A^\pm}  =\frac{e^{2\pi i (\Phi_B/\Phi_0)^\pm} r_\rightarrow^\pm}{1-e^{2\pi i (\Phi_B/\Phi_0)^\pm} t_\leftarrow^\pm}.\end{equation}
Here, we have included the superscript $\pm$ on $(\Phi_B/\Phi_0)$ as a reminder that we must use the appropriate version of Eq.~\eqref{global} corresponding to each sector.
The spin current evaluates to
\begin{equation}
\begin{split}(2/v_0)j_s = \psi(y)^\dagger\psi(y)= |A^+|^2 + |r_\rightarrow^+A^++t_\leftarrow^+B^+|^2  \\ + |A^-|^2 + |r_\rightarrow^-A^- + t_\leftarrow^- B^-|^2.\end{split} 
\end{equation}
Using Eq.~(\ref{Gammas}) to eliminate $B$'s, this expression becomes 
\begin{equation}
\begin{split} \label{js_simp}(2/v_0)j_s = |A^+|^2\left(1+\left|\frac{r_\rightarrow^+ }{1-e^{2\pi i (\Phi_B/\Phi_0)^+}   t_\leftarrow^+ }\right|^2\right) \\ +  |A^-|^2\left(1+\left|\frac{r_\rightarrow^- }{1-e^{2\pi i (\Phi_B/\Phi_0)^-}  t_\leftarrow^-}\right|^2\right) .\end{split} 
\end{equation}
From the constraint that we found on the magnitude of reflection coefficients we see that $|r_\rightarrow^+|^2=|r_\rightarrow^-|^2$. Furthermore,  from Eq.~(\ref{global}) we observe that $e^{2\pi i (\Phi_B/\Phi_0)^+} t_\leftarrow^+= e^{2\pi i (\Phi_B/\Phi_0)^-}  t_\leftarrow^-$ (again, since $\phi_{t^+}^\circ= \phi_{t^-}^\circ$ and that $|t|$'s are the same). As a result of this, the two quantities in parentheses in Eq.~\eqref{js_simp} are equal and can be factored out:
\begin{equation}
\label{js_simp}(2/v_0)j_s = (|A^+|^2+|A^-|^2)\left(1+\left|\frac{r_\rightarrow^+ }{1-e^{2 \pi 2\pi i (\Phi_B/\Phi_0)^+}  t_\leftarrow^+ }\right|^2\right).
\end{equation}
We can write a very similar expression for $j_p$ which differs from this expression only in a minus sign and also contains the same factor $|A^+|^2+|A^-|^2$. Therefore in writing $j_s/j_p$ those terms cancel out and we arrive at Eq.~(\ref{jsp1}).



\end{document}